\pdfoutput=1
\documentclass[11pt,twoside,a4paper,cmspaper,final,collab]{cms-tdr}
\def\svnVersion{227494c}\def\svnDate{2024/01/05}\def\cmsCernNoTag{CERN-EP-2023-064}\def\cmsCernDate{\today}\def\cmsMessage{Published in the Journal of High Energy Physics as \href{http://dx.doi.org/10.1007/JHEP12(2023)189}{\doi{10.1007/JHEP12(2023)189}.}}
\begin{document}\cmsNoteHeader{EXO-20-012}

\DeclareRobustCommand{\PQqst}{{\HepParticle{q}{}{\ast}}\Xspace}
\DeclareRobustCommand{\PAQqst}{{\HepAntiParticle{q}{}{\ast}}\Xspace}
\DeclareRobustCommand{\PQbst}{{\HepParticle{b}{}{\ast}}\Xspace}
\DeclareRobustCommand{\PAQbst}{{\HepAntiParticle{b}{}{\ast}}\Xspace}

\newcommand{\McPqst}{\ensuremath{M_{\PQqst}}\xspace}
\newcommand{\McPbst}{\ensuremath{M_{\PQbst}}\xspace}

\newcommand{\Mth}{\ensuremath{M_\text{th}}\xspace}
\newcommand{\sigmaBR}{\ensuremath{\sigma\mathcal{B}}\xspace}
\newcommand{\sqrts}{\ensuremath{\sqrt{s}}\xspace}
\newcommand{\sqrtsXIII}{\ensuremath{\sqrts = 13\TeV}\xspace}
\newcommand{\gammajet}{\ensuremath{\PGg \text{+jet}}\xspace}
\newcommand{\gammabjet}{\ensuremath{\PGg \text{+}\PQb \text{ jet}}\xspace}
\newcommand{\Zjet}{\ensuremath{\PZ \text{+jet}}\xspace}
\newcommand{\Wjet}{\ensuremath{\PW \text{+jet}}\xspace}

\cmsNoteHeader{EXO-20-012}
\title{Search for resonances in events with photon and jet final states in proton-proton collisions at \texorpdfstring{\sqrtsXIII}{sqrt(s) = 13 TeV}}
\author{The CMS Collaboration}
\date{\today}

\abstract{
A search for resonances in events with the \gammajet final state has been performed using proton-proton collision data collected at \sqrtsXIII by the CMS experiment at the LHC. The total data analyzed correspond to an integrated luminosity of 138\fbinv. Models of excited quarks and quantum black holes are considered. Using a wide-jet reconstruction for the candidate jet, the \gammajet invariant mass spectrum measured in data is examined for the presence of resonances over the standard model continuum background. The background is estimated by fitting this mass distribution with a functional form. The data exhibit no statistically significant deviations from the expected standard model background. Exclusion limits at 95\% confidence level on the resonance mass and other parameters are set. Excited light-flavor quarks (excited bottom quarks) are excluded up to a mass of 6.0 (3.8)\TeV. Quantum black hole production is excluded for masses up to 7.5 (5.2)\TeV in the Arkani-Hamed--Dimopoulos--Dvali (Randall--Sundrum) model. These lower mass bounds are the most stringent to date among those obtained in the \gammajet final state.
}
\hypersetup{
pdfauthor={CMS Collaboration},
pdftitle={Search for resonances in events with photon and jet final state in proton-proton collisions at sqrt(s) = 13 TeV},
pdfsubject={CMS},
pdfkeywords={CMS, beyond standard model, BSM, resonances}}

\maketitle

\section{Introduction}

Various theoretical proposals of beyond the standard model (BSM) physics predict the production of new resonances at hadron collider experiments. In the last decade, experiments at the CERN LHC have searched extensively for 
signatures of such phenomena to help understand some of the unresolved questions concerning features of the standard model (SM) of particle physics, for example the existence of only three fermion families and the hierarchy problem~\cite{hierarchy}. A variety of models have been proposed to  address these questions, some of which predict new resonances producing a \gammajet final state. Models of compositeness~\cite{compositeness_theories2, Baur:1987ga, Baur1989kv} address the existence of three fermion families, and other models address the hierarchy problem by predicting the existence of extra spatial dimensions~\cite{ArkaniHamed:1998rs,Randall:1999ee}.

The composite models postulate that the quarks and leptons are composed of more fundamental particles, known as ``preons''~\cite{Eichten:1984eu}. These preons experience a new force that becomes very strong at the compositeness scale ($\Lambda$), when forming the quark and lepton composite states. This paper considers the excited states of composite quarks produced via gauge-mediated interactions with the dominant process $\PQq \Pg \to \PQqst$, and decaying to SM particles. It is assumed that the compositeness scale $\Lambda$ is equal to the mass of the excited quark \McPqst.

The interactions between excited quarks, SM quarks, and the gauge bosons can be described by an effective Lagrangian~\cite{Bhattacharya:2009xg}
\begin{linenomath}
\begin{equation}
\mathcal{L}_\text{eff} = \frac{1}{2 \, \Lambda}\PAQqst \, \sigma^{\mu\nu}
\left[g_{s}f_{s}\frac{\lambda_{a}}{2}F^{a}_{\mu\nu}\;+\;gf\frac{\tau}{2}W_{\mu\nu}\;+\;g'f'\frac{Y}{2}B_{\mu\nu} \right] q + \text{h.c.}, 
\label{eq:lagrangian}
\end{equation}
\end{linenomath}
where $\PAQqst$ is the right-handed excited quark field and $\PQq$ is the left-handed quark field; $F^{a}_{\mu\nu}$, $W_{\mu\nu}$, and $B_{\mu\nu}$ are the field strength tensors associated with the SU(3), SU(2), and U(1) gauge fields, respectively; $\lambda_{a}$, $\tau$, and $Y$ are the corresponding Lie algebra generators; $g_{s}$, $g$, and $g'$ are the gauge couplings; $\sigma$ is the Pauli spin matrix; and $f_{s}$, $f$, and $f'$ are coupling strengths between the excited quark and SM quarks. An assumption is also made that $f_s$, $f$, and $f'$ have identical values and henceforth these coupling strengths will be collectively referred to as $f$.

\begin{figure}[h]
\centering
\includegraphics[width=0.49\textwidth]{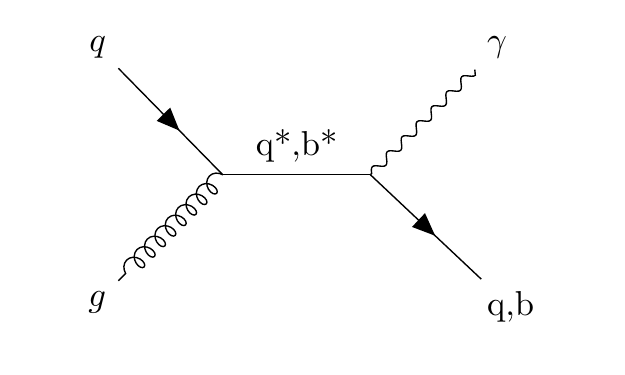}
\includegraphics[width=0.49\textwidth]{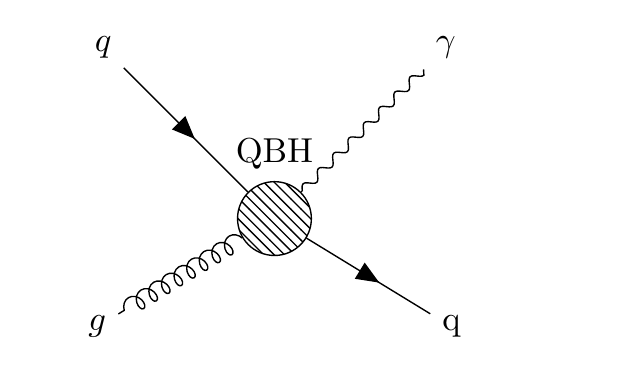}
\caption{Illustrative Feynman diagrams for $\PQq\Pg\to\gammajet$ resonance signal models of \PQqst, \PQbst (left), and  QBH production (right).}
\label{fig:feynman}
\end{figure}

In this paper, we consider excited light-flavor quarks and bottom quarks (\PQqst and \PQbst, respectively) radiating a photon and thus resulting in the \gammajet final state. The corresponding Feynman diagram for the production and decay of the \PQqst, \PQbst signals through gauge mediation is shown in Fig.~\ref{fig:feynman} (left). Additionally, an interpretation for \PQbst production from contact interactions (CI)~\cite{Baur1989kv} is also considered for the search of excited bottom quarks. Similar searches for excited quarks have been performed previously by the CMS~\cite{Khachatryan:2014aka, Khachatryan:2015sja, Khachatryan:2015dcf,dijet,Sirunyan:2017fho,CMS:2022zoc} and ATLAS~\cite{Aad:2015ywd,atlas_qbh,ATLAS:2018tfk,ATLAS:2019fgd} Collaborations at the LHC and at other high-energy particle colliders~\cite{Abe:1993sv, Abe:1997hm,Aaltonen:2008dn,D0:2003lui} and have reported no evidence of a signal. Lately, the ATLAS and CMS experiments have set exclusion limits of 5.3~\cite{atlas_qbh} and 5.5\TeV~\cite{Sirunyan:2017fho} on the mass of \PQqst in the \gammajet channel with data corresponding to 36 and 37\fbinv of integrated luminosity, respectively, collected at \sqrtsXIII. In the same run, the CMS experiment also has excluded the \PQbst in the \gammajet channel up to a mass of 1.8\TeV.

In addition, we also consider the models of Arkani-Hamed, Dimopoulos, and Dvali (ADD)~\cite{ArkaniHamed:1998rs}, and Randall--Sundrum (RS1)~\cite{Randall:1999ee}, which have extra spatial dimensions. The basic assumption of these models is that the fundamental gravitational scale (\MD) can be as low as 1\TeV, while the size of the extra dimensions can be as large as 1 millimeter, assuming the extra dimensions are only accessible to gravity and not
to the SM fields. The ADD model assumes ($4+n$) dimensions with $n=6$ flat compact extra dimensions, and RS1 considers $n=1$ extra dimension with a warped metric. As a consequence, the models predict the production of microscopic black holes~\cite{Dimopoulos:2001hw}, which can be either semi-classical black holes or quantum black holes (QBHs)~\cite{Calmet:2008dg,Meade:2007sz}. In this model, the threshold mass \Mth for production of a QBH at the LHC is assumed to be \MD, and the maximum mass limit is set to the smaller of either \sqrtsXIII or 3\MD, the mass beyond which the quantum description of the black hole does not hold. A QBH with a mass in this range will decay preferentially to a few-particle state, rather than undergo thermal decay. The QBH production occurs with a continuous QBH mass spectrum, falling rapidly from \Mth toward the maximum mass limit; however, because of radiation and experimental effects on the low-mass side of the mass spectrum, the signal shape resembles a resonance-like structure, as shown in Fig.~\ref{fig:Signal_shapes}. Previous searches for QBHs performed at \sqrtsXIII by the CMS and ATLAS experiments are reported in Refs.~\cite{dijet,Khachatryan:2015sja,QBH,CMS:2017caz,CMS:2018ucw,atlas_qbh,ATLAS:2017eqx}. This paper also reports the first search by the CMS experiment for QBHs in the \gammajet final state (as shown in the Feynman diagram of Fig.~\ref{fig:feynman} (right)).
The ATLAS Collaboration has excluded QBH production in the \gammajet channel with mass up to 7.1 (4.4)\TeV in the ADD (RS1) model using 13\TeV data corresponding to 37\fbinv of integrated luminosity~\cite{atlas_qbh}.

The main backgrounds in this analysis are quantum chromodynamics (QCD) initiated \gammajet production and multijet production, and electroweak \ensuremath{\PW/\PZ \text{+jet}} processes of the SM. The \gammajet background is an irreducible background, and is the largest contributing. Despite its large cross section, the QCD multijet background is only the second largest, contributing when one of the jets is rich in electromagnetic energy and consequently misidentified as a photon. This background also includes events in which a jet fragments into an energetic \Pgpz or \Pgh meson decaying into two overlapping photons that are reconstructed as a single photon in the detector. Electroweak processes mainly contribute via the production of \ensuremath{\PW/\PZ \text{+jet}}, with the heavy boson decaying leptonically, following the scenario where an electron mimics a photon because of missing hits in the tracker, and the jet passes the candidate selection.

This search for excited quarks and QBHs is performed using proton-proton ($\Pp\Pp$) collision data collected at \sqrtsXIII by the CMS experiment, corresponding to an integrated luminosity of 138\fbinv. The analysis strategy consists of searching for a resonance-like excess in the \gammajet invariant mass, above the expected SM background. In order to improve the mass resolution for resonances decaying to the \gammajet final state, wide jets are reconstructed from nominal-size jets to capture final-state radiation (FSR)~\cite{Khachatryan:2015sja,Khachatryan:2015dcf,dijet,CMS:2022zoc}. The background is estimated directly from data. However, MC-simulated background samples, which have been validated against data, are used to perform optimizations for the event selection. This analysis reaches higher sensitivity as compared to the past CMS search~\cite{Sirunyan:2017fho} and previous similar searches because of the improved \PQb jet tagging efficiency, and better signal mass resolution achieved through the use of wide jets. In addition, the integrated luminosity is approximately four times that used in earlier searches.   

This paper is organized as follows: Section~\ref{sec:detector} provides a brief description of the CMS detector. The triggers used for data selection are discussed in Section~\ref{sec:trigger}. In Section~\ref{sec:samples}, the data sets and simulation samples used in this analysis are described. The event reconstruction and selection criteria are detailed in Section~\ref{sec:selection}. The background estimation and systematic uncertainties are discussed in Sections~\ref{sec:modeling} and~\ref{sec:systematic}, respectively. The results and their interpretation are presented in Section~\ref{sec:results}, followed by a summary given in Section~\ref{sec:summary}. Tabulated results are provided in the HEPData record for this analysis~\cite{hepdata}.

\section {The CMS detector}
\label{sec:detector}
The central feature of the CMS apparatus is a superconducting solenoid of 6\unit{m} internal diameter, providing a magnetic field of 3.8\unit{T}. Within the solenoid volume are a silicon pixel and strip tracker, a lead tungstate crystal electromagnetic calorimeter (ECAL), and a brass and scintillator hadron calorimeter (HCAL), each composed of a barrel and two endcap sections. Forward calorimeters extend the pseudorapidity ($\eta$) coverage provided by the barrel and endcap detectors. The muon detection system consists of a gas-ionization detector installed outside the solenoid and embedded in the steel flux-return yoke.

In the barrel section of the ECAL, an energy resolution of about 1\% is achieved for unconverted or late-converting photons in the tens of \GeVns energy range. The energy resolution of the remaining barrel photons is about 1.3\% up to $\abs{\eta} = 1$, changing to about 2.5\% at $\abs{\eta} = 1.4$. In the endcaps, the energy resolution is about 2.5\% for unconverted or late-converting photons, and between 3 and 4\% for the other ones~\cite{CMS:EGM-14-001}. The jet energy resolution amounts typically to 15--20\% at 30\GeV, 10\% at 100\GeV, and 5\% at 1\TeV~\cite{Khachatryan:2016kdb}. A more detailed description of the CMS detector, together with a definition of the coordinate system used and the relevant kinematic variables, can be found in Ref.~\cite{Chatrchyan:2008zzk}.

Events of interest are selected using a two-tiered trigger system. The first level, composed of custom hardware processors, uses information from the calorimeters and muon detectors to select events at a rate of around 100\unit{kHz} within a fixed latency of about 4\mus~\cite{Sirunyan:2020zal}. The second level, known as the high-level trigger (HLT), consists of a farm of processors running a version of the full event reconstruction software optimized for fast processing, and reduces the event rate to around 1\unit{kHz} before data storage~\cite{Khachatryan:2016bia}.

\section {Trigger selection}
\label{sec:trigger}
Events used in this analysis must pass a HLT single-photon trigger, which has a photon transverse momentum \pt threshold of 175\GeV in 2016, and 200\GeV in 2017--2018. The higher \pt threshold in 2017--2018 is set to optimize available bandwidth and storage of events for offline analysis. The trigger also requires that the fraction of energy associated with the photon deposited in the HCAL is less than 15\% of that deposited in the ECAL ($H/E < 0.15$). The efficiency of single-photon triggers is calculated using a single-photon data set with a low-\pt requirement and is found to be 99\%, above the photon \pt value of 240\GeV. To calculate the possible bias on single-photon trigger due to any systematic detector effects, the efficiency is also calculated using jet- and muon-triggered independent data sets. The trigger efficiency with the independent data sets is found to be about 97--98\% and the trigger inefficiency is taken as an estimate of systematic uncertainty.

\section {Simulated data samples}
\label{sec:samples}
The signal and background samples are obtained from event generators using Monte Carlo (MC) simulations. The generated events are processed through the full CMS detector simulation based on \GEANTfour~\cite{Agostinelli:2002hh}. Additional pp interactions in the same and nearby proton bunch crossings, referred to as pileup, are also simulated and a weight is applied to simulated events to reproduce the same pileup distribution as measured in data. The simulation uses the CUETP8M1 and CP5 underlying event tunes~\cite{Khachatryan:2015pea,Sirunyan:2019dfx} for the 2016 and 2017--2018 data sets, respectively. The samples with the CUETP8M1 tune use the NNPDF3.0 LO~\cite{NNPDF:2014otw} parton distribution functions (PDFs), while simulated samples generated with the CP5 tune use the NNPDF3.1 next-to-next-LO~\cite{Ball:2017nwa} PDFs, unless otherwise mentioned.

The signal samples for \PQqst and \PQbst are generated at leading order (LO) with the \PYTHIA 8.205 (8.230) event generator~\cite{Sjostrand:2014zea} for 2016 (2017/2018) using coupling values $f = 1.0$, 0.5, and 0.1. These \PQqst and \PQbst signal events are generated for masses in the range from 1.0 to 7.0\TeV at intervals of 1.0 and 0.5\TeV, respectively. For \PQbst signals, only events with \McPbst within $\pm$20\% of the input \McPbst  are considered in order to avoid sizable off-shell nonresonant contributions, which are otherwise present for high-\McPbst scenarios with coupling parameters $f=0.5$ and 1.0 but do not contribute to the search sensitivity. The signals for intermediate mass values are extracted with linear interpolation at intervals of 0.1 TeV. This procedure is tested and provides similar shapes to those generated.

\begin{figure}[h]
\centering
\includegraphics[width=0.7\textwidth]{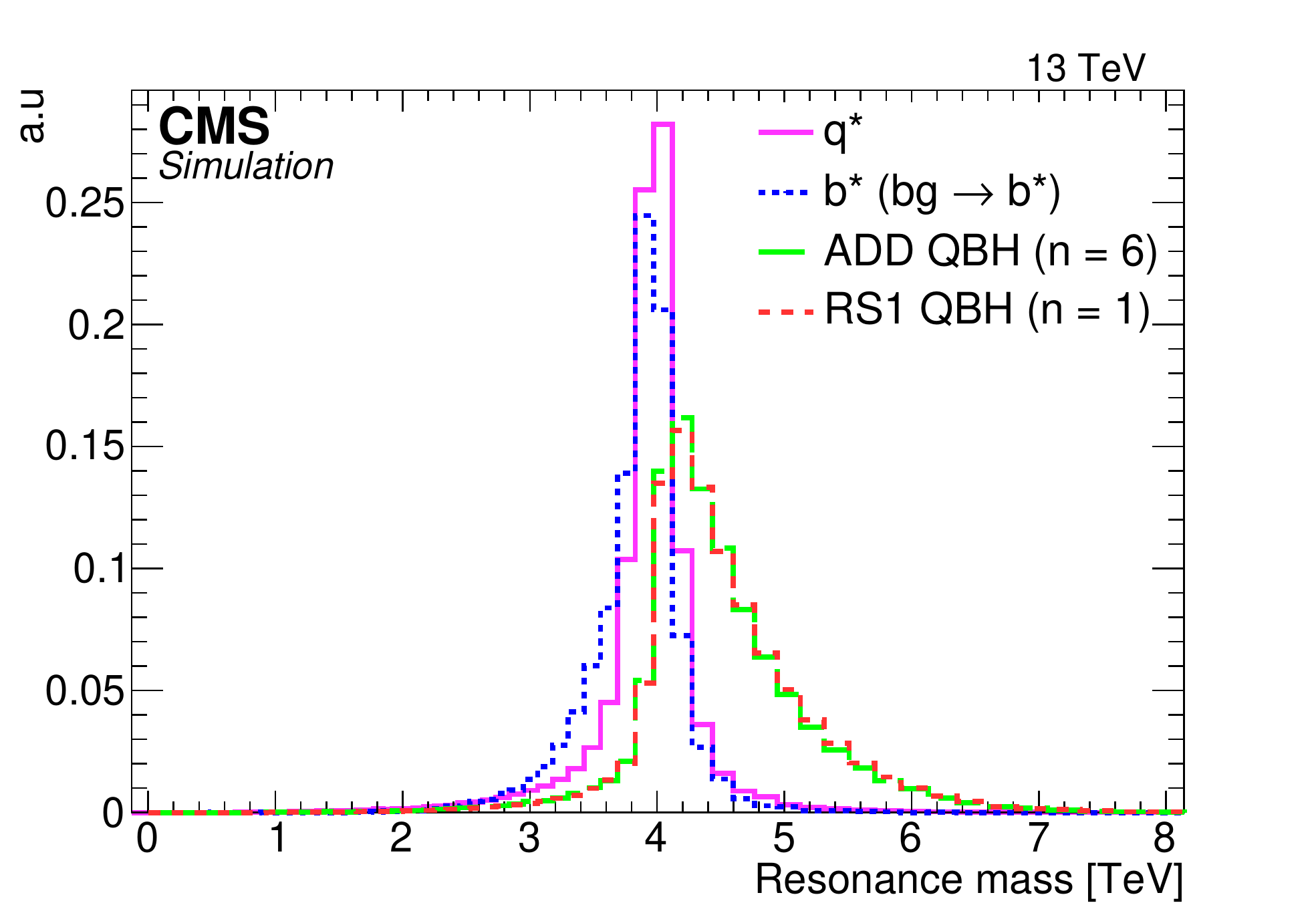}
\caption{Distributions of reconstructed \gammajet invariant mass for the q* and b* signal models with resonance mass of 4\TeV, and for QBH from the ADD and RS1 models with $\Mth=4$\TeV.}
\label{fig:Signal_shapes}
\end{figure}

The QBH signal events in the ADD ($n=6$) and the RS1 ($n=1$) models are generated with the QBH generator~\cite{Gingrich} version v3.0 using CTEQ6L~\cite{Pumplin:2002vw} PDFs. The signals are generated for masses ranging from 3.0 to 8.5\TeV for the ADD model and from 2.0 to 7.0\TeV for the RS1 model at intervals of 0.5\TeV. The \gammajet mass distribution for QBH is similar between the RS1 and ADD models; therefore, the RS1 simulated sample is used for the ADD model for $\Mth<3\TeV$. The resonance shapes for \PQqst, \PQbst, and QBH for the ADD ($n=6$) and RS1 ($n=1$) models for a mass of 4\TeV are shown in Fig.~\ref{fig:Signal_shapes}.

\begin{figure}[h]
\centering
\includegraphics[width=0.7\textwidth]{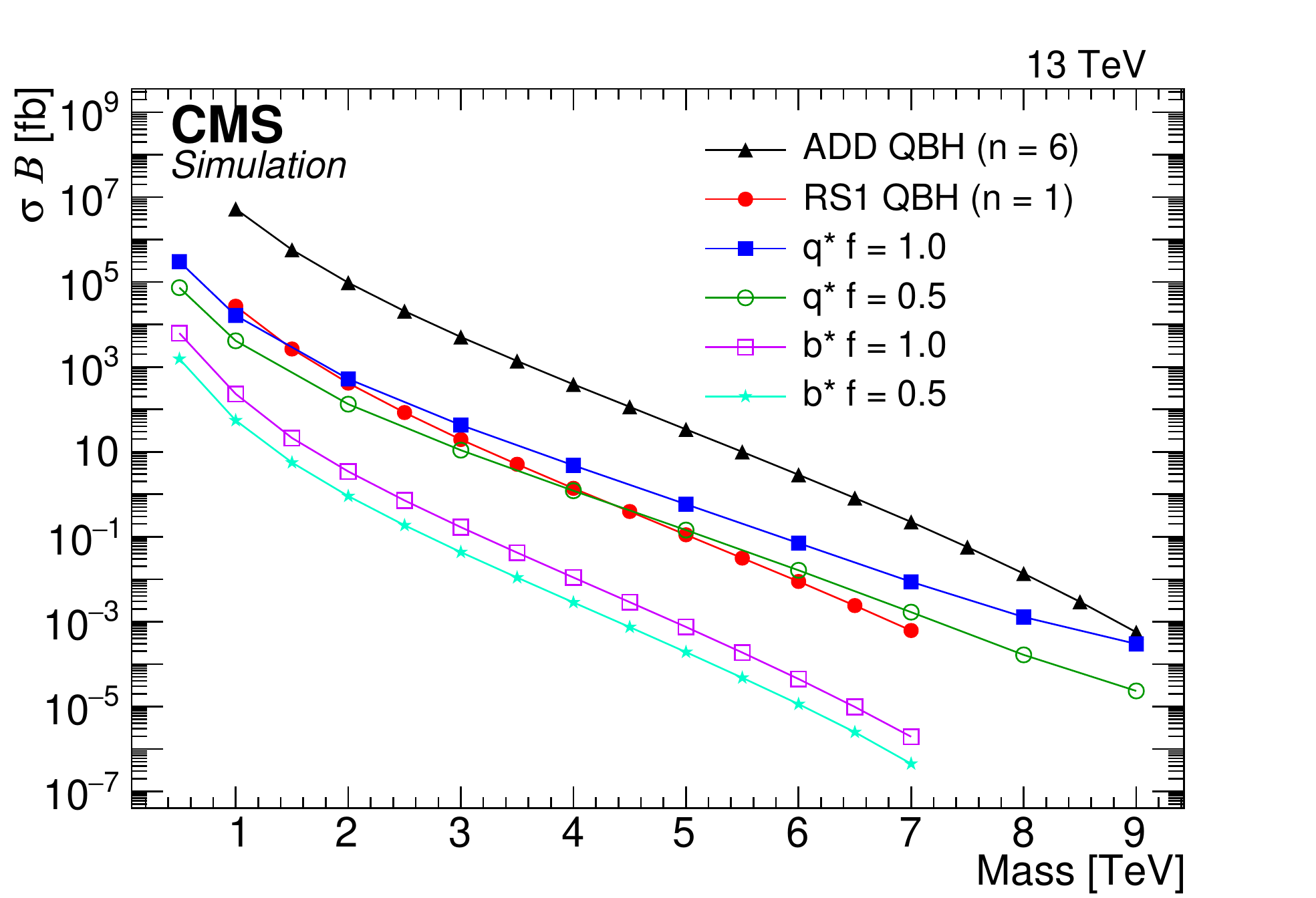}
\caption{The product of the cross section and branching fraction, \sigmaBR, as a function of the resonance mass of \PQqst or \PQbst signals (\McPqst or \McPbst) or the threshold mass of QBH signal models (\Mth).}
\label{fig:Signal_XS}
\end{figure}

Figure~\ref{fig:Signal_XS} shows the product of the cross section ($\sigma$) and branching fraction $\mathcal{B}$ for the production at \sqrtsXIII of the photon and jet final state as a function of the resonance mass, for excited quark models, or the threshold mass, for QBH models.

The \gammajet, Drell--Yan, and \Wjet backgrounds are simulated using the \MGvATNLO version 2.2.2 (2.4.2) generators for 2016 (2017--2018)~\cite{Alwall:2014hca} at LO in QCD with the MLM matching scheme~\cite{Alwall:2007fs} between jets produced via matrix element calculations and from parton showers. The QCD multijet background simulation is obtained at LO with the \PYTHIA generator using version 8.205 (8.230) for 2016 (2017--2018). Hadronization and parton showering of all simulated samples are performed using the \PYTHIA program versions 8.205 and 8.230 with the CUETP8M1 and CP5 tunes, respectively.

\section {Event reconstruction and selection}
\label{sec:selection}
The particle-flow (PF) algorithm~\cite{CMS:2017yfk} reconstructs and identifies photons, electrons, muons, and charged and neutral hadrons (called PF candidates), with an optimized combination of information from the various elements of the CMS detector. The energy of photons is obtained from the ECAL detector. The clusters of energy deposited in the ECAL are combined together within a defined geometric area to form a supercluster around the seed cluster~\cite{CMS:EGM-17-001}. The energy of electrons is determined from a combination of the electron momentum at the primary interaction vertex as determined by the tracker, the energy of the corresponding ECAL cluster, and the energy sum of all bremsstrahlung photons spatially compatible with their origin from the electron track. The energy of muons is obtained from the curvature of the corresponding track. The energy of charged hadrons is determined from a combination of their momentum measured in the tracker with the matched ECAL and HCAL energy deposits, corrected for the response function of the calorimeters to hadronic showers. Finally, the energy of neutral hadrons is obtained from the corresponding corrected ECAL and HCAL energies.

For each event, hadronic jets are reconstructed from PF candidates using the infrared- and collinear-safe anti-\kt algorithm~\cite{Cacciari:2008gp,Cacciari2011ma} with a distance parameter of 0.4. Jet momentum is determined as the vectorial sum of all particle momenta in the jet, and is found from simulation to be, on average, within 5--10\% of the true momentum over the whole \pt spectrum and detector acceptance. Pileup interactions can contribute additional tracks and calorimetric energy depositions to the jet momentum. To mitigate this effect, charged particles identified to be originating from pileup vertices are discarded and an offset correction is applied to correct for remaining contributions. Jet energy corrections are derived from simulation to bring the measured response of jets to that of particle level jets on average. In situ measurements of the momentum balance in dijet, \gammajet, \Zjet, and multijet events are used to account for any residual differences in the jet energy scale between data and simulation~\cite{Khachatryan:2016kdb}.

The primary vertex (PV) is taken to be the vertex corresponding to the hardest scattering in the event, evaluated using tracking information alone, as described in Section~9.4.1 of Ref.~\cite{CMS-TDR-15-02}.

Candidate events are required to have at least one photon with transverse momentum $\pt^{\Pgg}>240\GeV$ in the barrel region ($\abs{\eta^{\Pgg}} < 1.44$) of the detector. Above the $\pt^{\Pgg}$ threshold of 240\GeV, the trigger is 99\% efficient. The photon  with the highest \pt is  selected as the candidate photon in the event.
To reject jets misreconstructed as photons, candidate photons are required to satisfy 
identification and isolation requirements~\cite{CMS:EGM-17-001} having a selection efficiency of 80\% on signal-like events. An electron veto is applied to remove photon candidates in the cases where the candidate's supercluster is matched to a reconstructed electron's track that has no missing hits in the innermost tracker layers.

The observed difference between data and MC simulation in the efficiency for photon identification and isolation is accounted for by applying a data-to-simulation scale factor, as a function of \pt and $\eta$~\cite{CMS:EGM-17-001}.

Each candidate event is also required to have at least one jet that passes identification criteria in order to remove spurious jets arising from detector noise~\cite{CMS-PAS-JME-16-003}. Among the selected jets, a list of jets having at least $\pt^\text{jet} > 30\GeV$ and $\abs{\eta^\text{jet}} < 3.0$ is made in decreasing order of \pt. Setting a minimum value for $\pt^\text{jet}$ reduces the number of pileup jets in the selected events.

The jets passing all the above criteria and calibrated as discussed above are clustered to form wide jets. The clustering captures FSR and improves the \gammajet invariant mass resolution. To construct
wide jets, the highest \pt jet is used as a seed jet, and other jets are merged with it if they are within $\Delta R < 1.1$. The jets are merged by adding the four momenta of jets. The resultant jet is stored as a wide jet and the constituent jets are removed from the jet list. The procedure is repeated with the remaining highest \pt jet as a seed jet to form the next wide jet. These clustered wide jets are  ordered in \pt and the highest one is used as the final jet candidate to study signal-like characteristics, such as the invariant mass of the photon and jet. The use of wide jets improves the mass resolution in the $\PQqst/\PQbst$ signal by 1.0--1.3\% over the entire mass range considered in this analysis. The leading wide jet is required to have $\pt(\text{wide jet})>170\GeV$ with $\abs{\eta(\mbox{wide jet})}<2.4$. The final selected  photon and wide-jet candidates are required to be separated by $\Delta R > 1.1$.

A large reduction in the number of QCD multijet events is achieved by the requirements on photon isolation. To suppress this background further, an additional requirement of $\abs{\Delta\eta}<1.5$ is imposed on the photon and wide-jet system. This requirement mainly reduces QCD dijet events produced by $t$-channel processes, which have a relatively large spread in $\Delta\eta$. If a leading wide-jet (or photon) candidate fails any of the selections described above, the event is discarded. The events selected are assigned to the "inclusive category" for the \PQqst and QBH signal searches.

The \PQbst signal search separates the data sets into two categories based on the identification of jets originating from \PQb quarks (\PQb jets). The identification of \PQb jets is performed using the neural network-based \textsc{DeepJet} tagger~\cite{BTV-16-002,DeepJet_tagger} at an operating point with 75\% \PQb quark efficiency and 1\% light quarks misidentification probability. The two categories are the \PQb tag category (events with at least one \PQb-tagged jet) and the 0-\PQb tag category (events without any \PQb-tagged jet). The sensitivity of the \PQbst signal search is improved by combining the results of the 0-\PQb tag category and the \PQb tag category. This is done to mitigate the effect of a slightly reduced acceptance of the \PQb tag category at higher masses.

The wide jet, in the case of a \PQbst, is constructed with the same process described previously, but with an additional requirement that at least one of the constituent jets within $\Delta R < 1.1$
must pass the \PQb tag discriminator selection and fall into the \PQb tag category. This jet will be referred to as a \PQb-tagged jet. In a wide jet, there could be one, or more \PQb-tagged jets and also light jets. The simulated distributions in both \PQb tag and 0-\PQb tag categories are weighted using scale factors (SF) that account for differences between data and simulation in the \PQb tagging efficiencies.

\begin{figure}[h]
\centering
\includegraphics[width=0.49\textwidth]{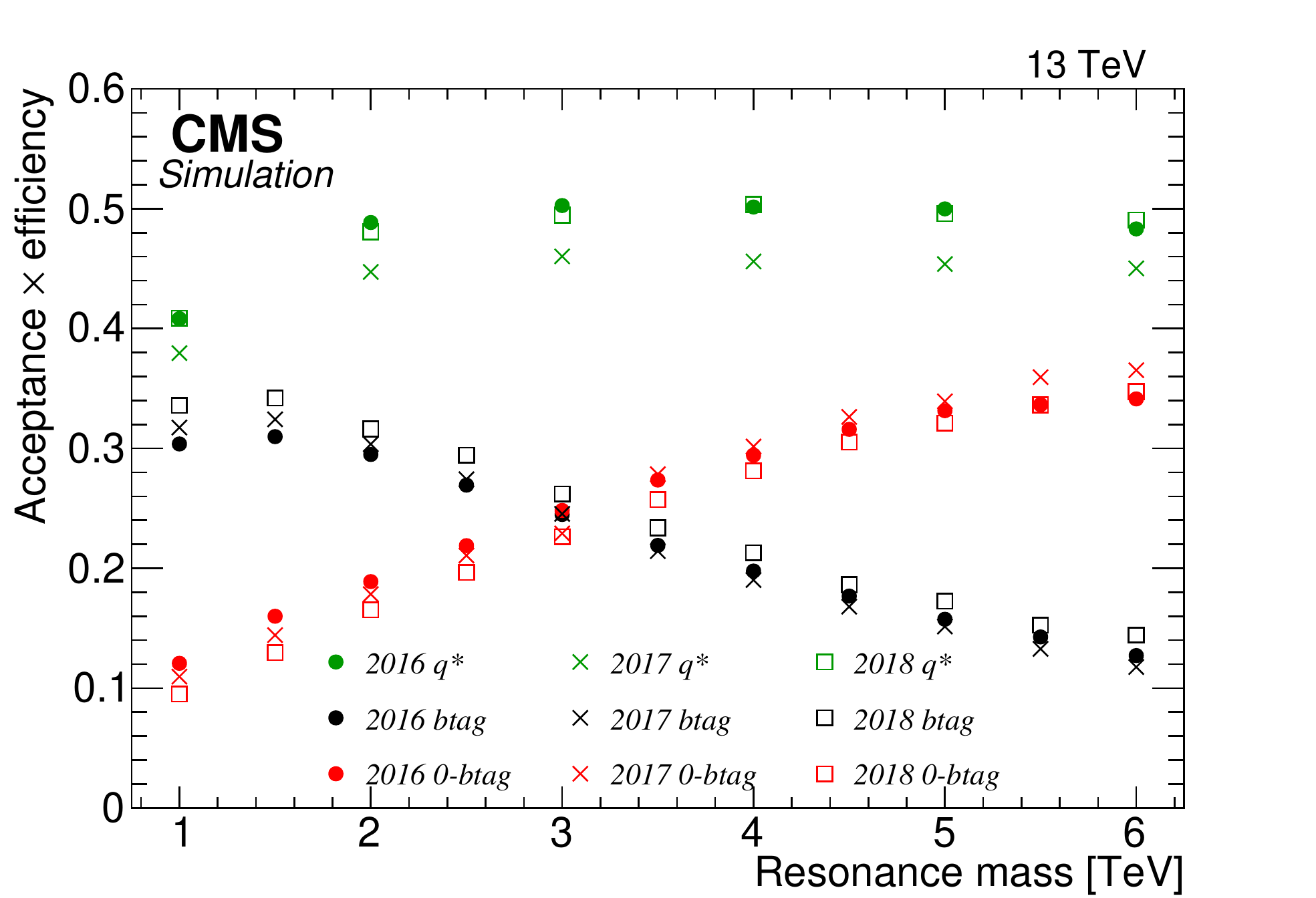}
\includegraphics[width=0.49\textwidth]{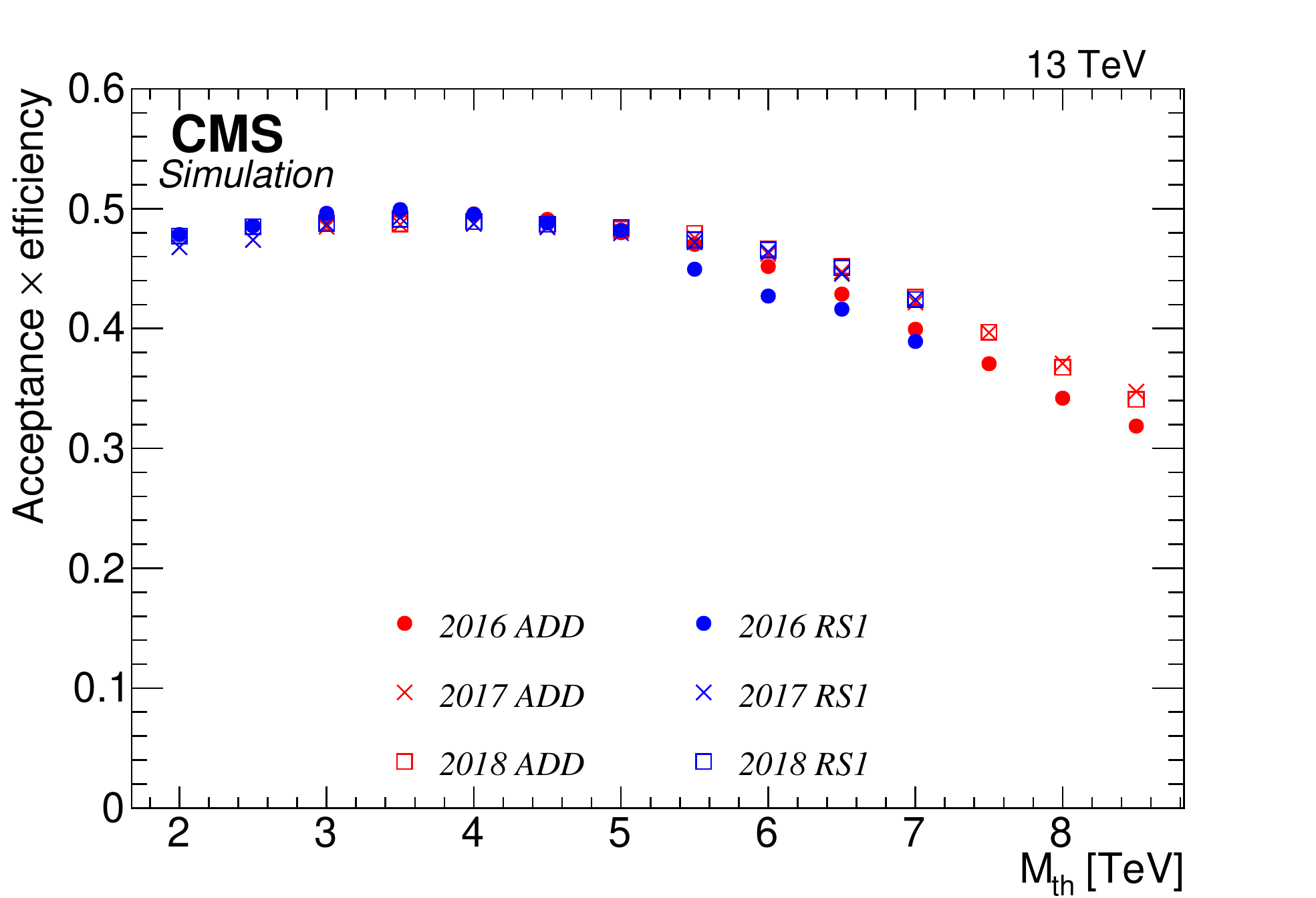}
\caption{The product of acceptance and efficiency for the \PQqst signal in the inclusive category and the \PQbst signal in the \PQb tag and 0-\PQb tag categories, for SM coupling $f = 1.0$ and for different resonance values (left), and the signal for the QBH (ADD/RS1) model in the inclusive category, for different \Mth values (right), for the years 2016, 2017, and 2018.}
\label{fig:acceff}
\end{figure}

To avoid the turn-on region in the photon + wide-jet invariant mass distribution that arises due to the various kinematic selections, only events with $m_{\gammajet}>0.76\TeV$ are considered, where \gammajet denotes the photon and wide-jet system. The products of acceptance and efficiency for the $\PQqst/\PQbst$ and QBH (ADD/RS1) signals are shown in Fig.~\ref{fig:acceff} for the years 2016, 2017, and 2018. For the ADD and RS1 signals with mass $\Mth > 5\TeV$, the reduction in the product of acceptance and efficiency mainly comes from the application of photon isolation criteria. However, the relatively small difference between the observed efficiencies across different years comes from the different pileup conditions and detector configurations, such as an additional pixel layer in the tracker~\cite{Phase1Pixel} and the upgraded HCAL~\cite{HCALtdrphase1upgrade}. In the following sections, we will refer to wide jets simply as jets.

\section {Signal and background modeling}
\label{sec:modeling}

To extract the background from data, the \gammajet mass spectrum is parameterized with the following empirical functional form
\begin{linenomath}
\begin{equation}
\frac{\rd{\sigma}}{\rd m} = \frac{P_{0}(1 - m/\sqrts)^{P_{1}}}{(m/\sqrts)^{P_{2} + P_{3} ln(m/\sqrts)}},
\label{eq:bkgfun}
\end{equation}
\end{linenomath}
where $P_{0}$, $P_{1}$, $P_{2}$, and $P_{3}$ are the fit parameters, and \sqrtsXIII. The search uses the signal plus background binned fit method, where the mass spectrum is fitted with this functional form, the normalization is allowed to float, and the signal templates are obtained from simulation. The parameters $P_{0}$, $P_{1}$, $P_{2}$, and $P_{3}$ in this functional form are kept floating. This same functional form has also been used in a number of earlier searches~\cite{Aaltonen:2008dn, Khachatryan:2014aka, Khachatryan:2015dcf, Aad:2015ywd, Sirunyan:2017fho} to describe the SM background distribution. The fitting range for the \PQqst, \PQbst, and RS1 signal is $0.76 < m_{\gammajet} < 8.40\TeV$, while for the ADD signal interpretation, where a better signal sensitivity is reached, it is extended up to 9.4\TeV.

\begin{figure}[h!]
\centering
\includegraphics[width=0.7\textwidth]{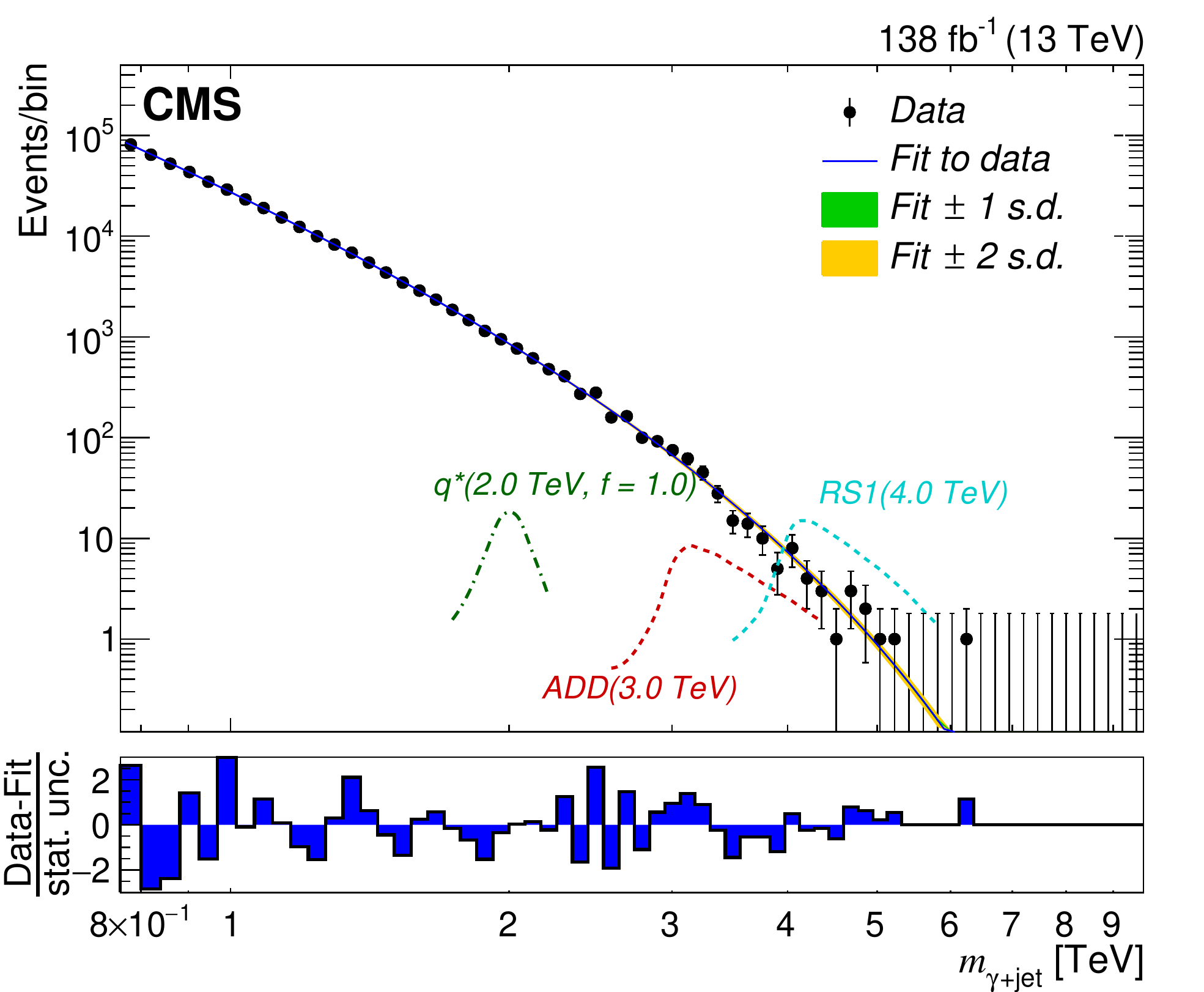}
\caption{Fit to the \gammajet invariant mass distribution in data with Eq.~\eqref{eq:bkgfun} for the "inclusive" category of \PQqst and QBH after all selections. Simulations of the \PQqst signal with a 2\TeV mass and coupling $f = 1.0$, the QBH ADD signal with $\Mth=3\TeV$, and the QBH RS1 signal with $\Mth=4\TeV$ are also shown. The signal distributions are normalized to their exclusion upper limit cross sections. The lower panel shows the difference between the data yield and the background prediction divided by the statistical uncertainty of the data. The green (inner) and yellow (outer) bands, respectively, represent the 68 and 95\% confidence level statistical uncertainties in the fit.}
\label{fig:MassFitq}
\end{figure}

\begin{figure}[h!]
\centering
\includegraphics[width=0.49\textwidth]{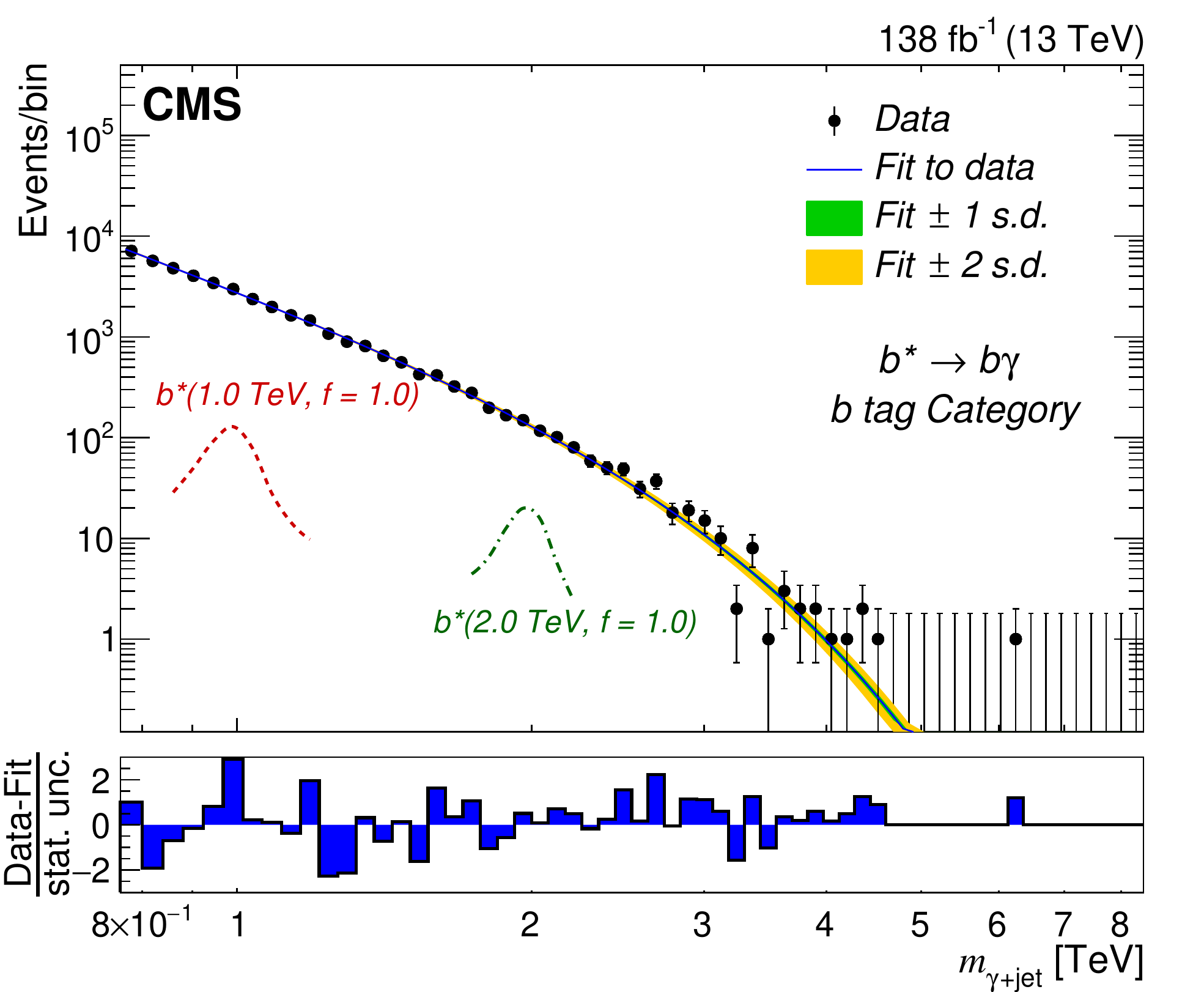}
\includegraphics[width=0.49\textwidth]{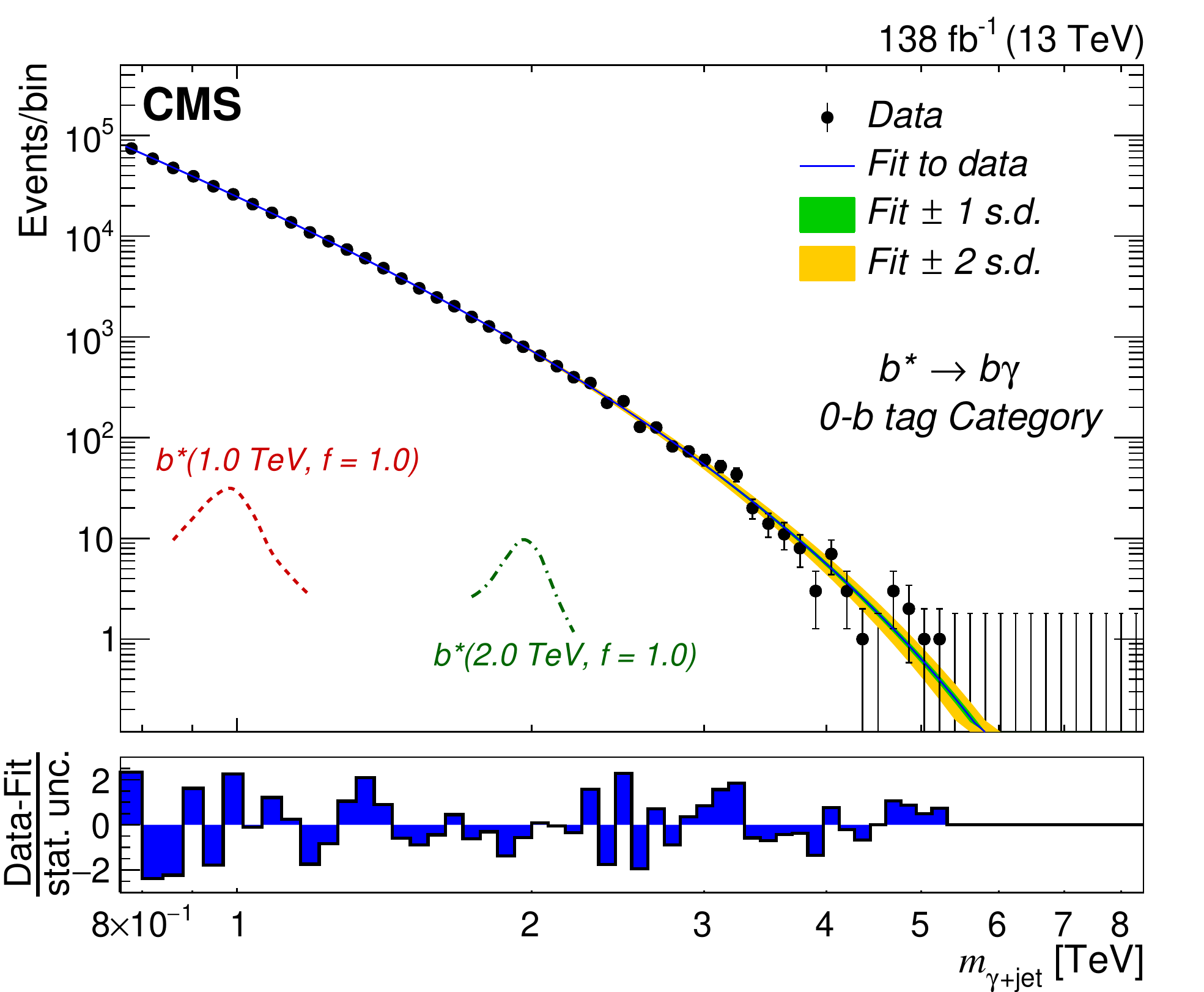}
\caption{Fits to the \gammajet invariant mass distributions in data with Eq.~\eqref{eq:bkgfun} for the \PQbst selection for the \PQb tag category (left) and 0-\PQb tag category (right). Simulations of \PQbst signals are shown for mass values of 1.0 and 2.0\TeV and for $f = 1.0$. The signal distributions are normalized to their exclusion upper limit cross sections. The lower panel shows the difference between the data yield and background prediction divided by the statistical uncertainty of the data. The green (inner) and yellow (outer) bands, respectively, represent the 68 and 95\% confidence level statistical uncertainties in the fit.}
\label{fig:MassFit}
\end{figure}

The invariant mass distributions for the \gammajet and \gammabjet channels, after all selections, are shown in Figs.~\ref{fig:MassFitq} and \ref{fig:MassFit}, respectively. The background-only fits performed on data using Eq.~\eqref{eq:bkgfun} and the expected signal resonances are also shown. The bin size of the mass distribution is varied to match the \gammajet mass resolution, which is approximately 4.5\% at 1\TeV and 3.3\% at 6\TeV.

Alternative polynomial functions with 4 and 5 parameters are also considered, to evaluate any possible bias due to
the chosen functional form shown in Eq.~\eqref{eq:bkgfun}. The choice of the background function is made on the basis of the results of the Fisher test~\cite{10.2307/2340521}. In order to evaluate the possible bias introduced by the specific choice of fit function, pseudo-experiment data sets are generated for a variety of functions used to represent the simulated SM background distributions, with each data set including an injected signal at the expected 95\% confidence level. These pseudo-experiment mass distributions are then fit with Eq.~\eqref{eq:bkgfun} to extract measured signal cross sections. The same procedure is repeated based on the measured mass distributions instead of on the fully simulated MC background distributions. In both cases, distributions of pull, defined as a difference between the injected and extracted signal cross sections divided by the statistical uncertainty of the extracted signal cross section, have mean biases of less than half a standard deviation. A bias of this magnitude is considered acceptable since it contributes at most about 11\% to the total uncertainty when added in quadrature with the statistical uncertainty. Following the practice of earlier searches~\cite{dijet,Sirunyan:2017fho,CMS:2022zoc}, in this search the statistical uncertainty of the fit is the only uncertainty considered in the background prediction. As a check, to estimate the effect of the background modeling bias, we included in the signal yield calculation the cross section uncertainty corresponding to the maximum potential bias and found that the effect of such a bias on the limit was negligible.

\section {Systematic uncertainties}
\label{sec:systematic}
The systematic uncertainties due to various sources can affect both the normalization and the shape of the
mass distribution. The dominant source of uncertainty comes from the parametrization of the background fit to data. The background uncertainties are considered as floating nuisance parameters, while the uncertainties in the signal predictions are implemented either as log-normal or Gaussian constraints on the nuisance parameters.

The signal yield and shape are affected by the uncertainties related to the energy or momentum measurements for the reconstructed photons and jets.

To evaluate the effect of uncertainties in wide jets, the jet-related uncertainties are first
applied to the constituent jets. The jet energy scale and jet energy resolution effects are measured by
applying the respective $\pm$1 standard deviation uncertainties in the four-momenta of the constituent jets and
reconstructing the mass of the \gammajet system. The effects of the jet energy scale and resolution on the mean and width of the signal shape are found to be in the range 1.0--2.0\%.

The photon energy scale and photon energy resolution are measured with $\PZ \to \Pep\Pem$ events, with the electrons reconstructed as photons, using the procedure described in Ref.~\cite{CMS:EGM-17-001}. The uncertainties in the photon energy scale and photon energy resolution translate into a variation of
1.0--1.5\% in the signal mean and width. The effects of uncertainties associated with the photon identification and isolation scale factors are found to be negligible for photons with \pt up to 500 GeV; however, beyond this \pt value, the associated uncertainty increases because of the extrapolation in its measurement at high \pt, and is taken to be 10\%. This uncertainty covers all related systematic effects in the procedure used in its measurement.

The \PQbst signal shape mean and yield have 0.2--3.0\% and 7.9--14.3\% uncertainties, respectively, obtained by
varying the \PQb tagging SF within its $\pm$1 standard deviation uncertainty. The other sub-dominant uncertainties for the \PQbst signal shape and normalization are found to be of the order of 1.0--2.0\%, as in the case of
\PQqst.

To account for the uncertainty in the HLT inefficiency as discussed in Section~\ref{sec:trigger}, a 2.5\% systematic uncertainty is used.

The uncertainties associated with the integrated luminosity measurement~\cite{CMS:lumi2016,CMS:lumi2017,CMS:lumi2018} and pileup are also considered. The integrated luminosities of the 2016, 2017, and 2018 data-taking periods are individually known with uncertainties in the 1.2--2.5\% range, while the total 2016--2018 integrated luminosity has an uncertainty of 1.6\%, used in this paper. The uncertainty due to pileup is calculated by varying the inelastic cross section of 69.4\unit{mb}~\cite{Sirunyan:2018nqx} by $\pm$5\%  and its impact is estimated to be $\sim$0.1\% on the signal yield. The uncertainty in the signal shapes due to interpolation for intermediate masses is considered to be $\sim$1\%, evaluated by comparing the interpolated mass shape with the corresponding generated mass point.

The statistical interpretation considers the systematic uncertainties to be fully correlated among different data-taking years, except for the integrated luminosity uncertainty, which is partially correlated between the years. The effect of uncorrelated systematic uncertainties is also studied by taking the respective
uncertainties for each data set separately and obtaining the results again. The results obtained are
found to be similar to the case when fully correlated systematic uncertainties are considered for all years.

The details of systematic uncertainties and their relative effects on the shape and normalization are listed in Table~\ref{Table:syst}.

\begin{table}[h]
\centering
\topcaption{Systematic uncertainties in the signal yields for the mass range analyzed.}
\begin{tabular}{l c c c}
Source                       & Type           & Uncertainty           & Applied to (\PQqst/\PQbst/QBH) \\
\hline
Integrated luminosity        & Normalization  & 1.6\%                    & All\\
Jet energy scale             & Shape (mean)   & $\sim$1.0\%              & All\\
Jet energy resolution        & Shape (width)  & 0.6--1.9\%               & All\\
Photon energy scale          & Shape (mean)   & $\sim$1.0\%              & All\\
Photon energy resolution     & Shape  (width) & 0.4--1.2\%               & All\\
Pileup                       & Normalization  & 0.1\%                    & All\\
High-\pt photon efficiency   & Normalization  & 10.0\%             & All\\
HLT efficiency               & Normalization  & 2.5\%                    & All\\
Signal shape interpolation   & Shape (mean)   & $\sim$1.0\%              & All\\
\PQb tag SF uncertainty      & Shape (mean)   & 0.2--3.0\%               & Only \PQbst \\
\PQb tag SF uncertainty      & Normalization  & 7.9--14.3\%              & Only \PQbst
\end{tabular}
\label{Table:syst}
\end{table}

\section {Results and interpretation}
\label{sec:results}
\begin{figure}[!htbp]
\centering
\includegraphics[width=0.49\textwidth]{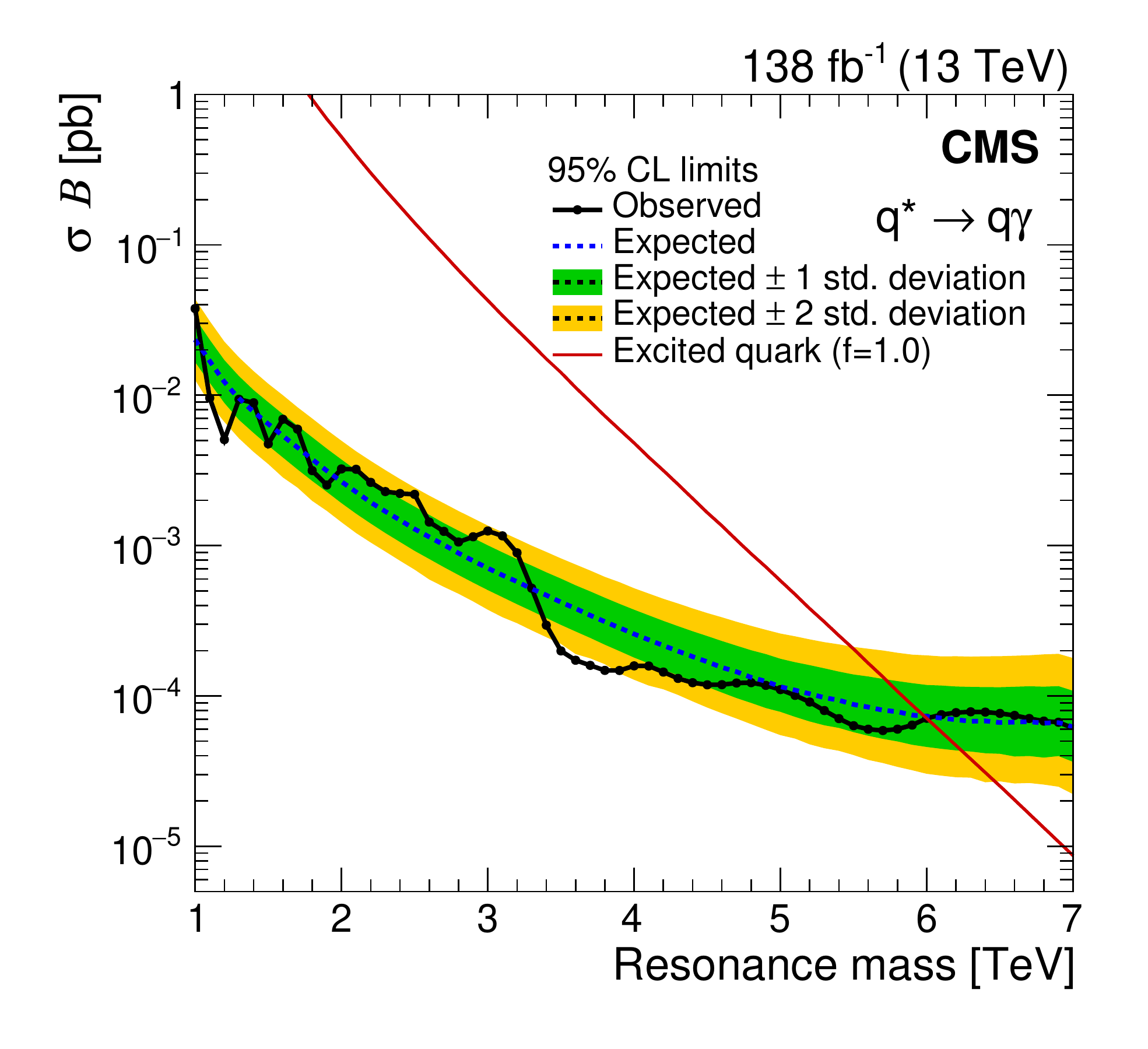}
\includegraphics[width=0.49\textwidth]{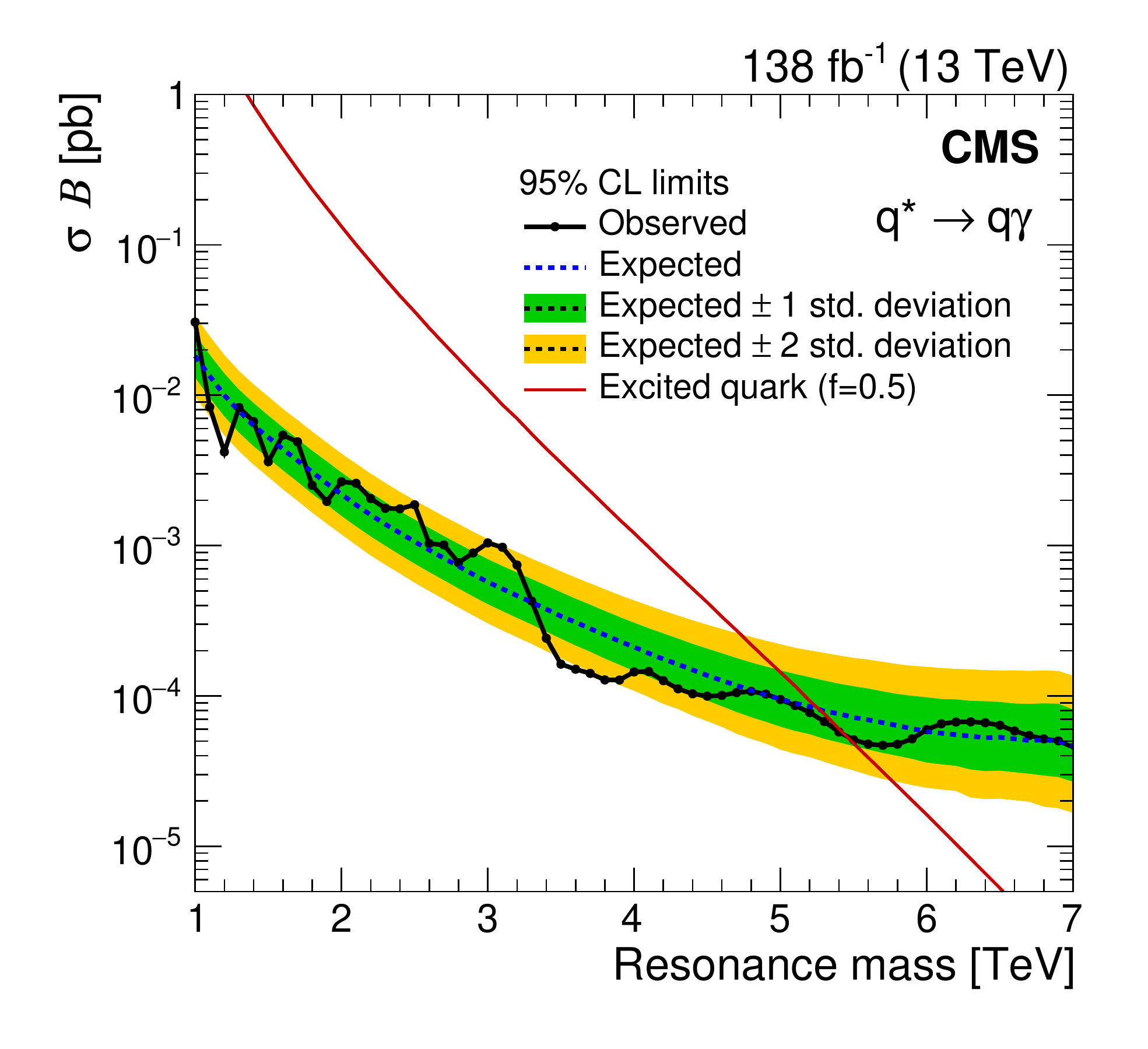}
\includegraphics[width=0.49\textwidth]{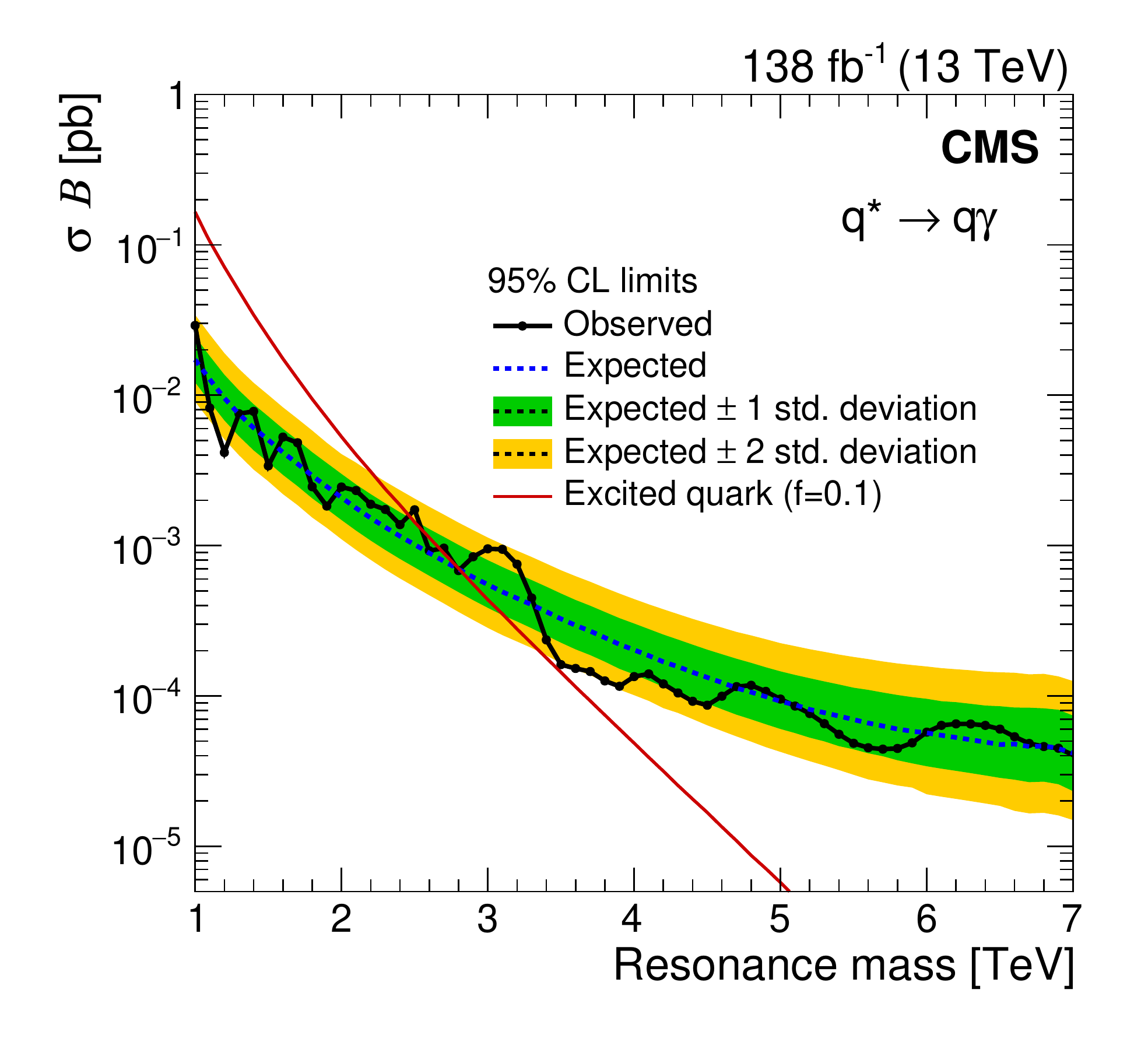}
\caption{The expected (dashed) and observed (solid) 95\% \CL upper limits on the product of the cross section and branching fraction, as functions of \PQqst mass, for coupling strength $f=1.0$ (upper left), $f=0.5$ (upper right), and $f=0.1$ (lower). The green (inner) and yellow (outer) bands correspond to 1 and 2 standard deviation uncertainties in the expected limit, respectively. The limits are compared with the theoretical predictions for \PQqst production for the corresponding coupling strengths.}
\label{fig:qstarlimit}
\end{figure}

\begin{figure}[!htb]
\centering
\includegraphics[width=0.49\textwidth]{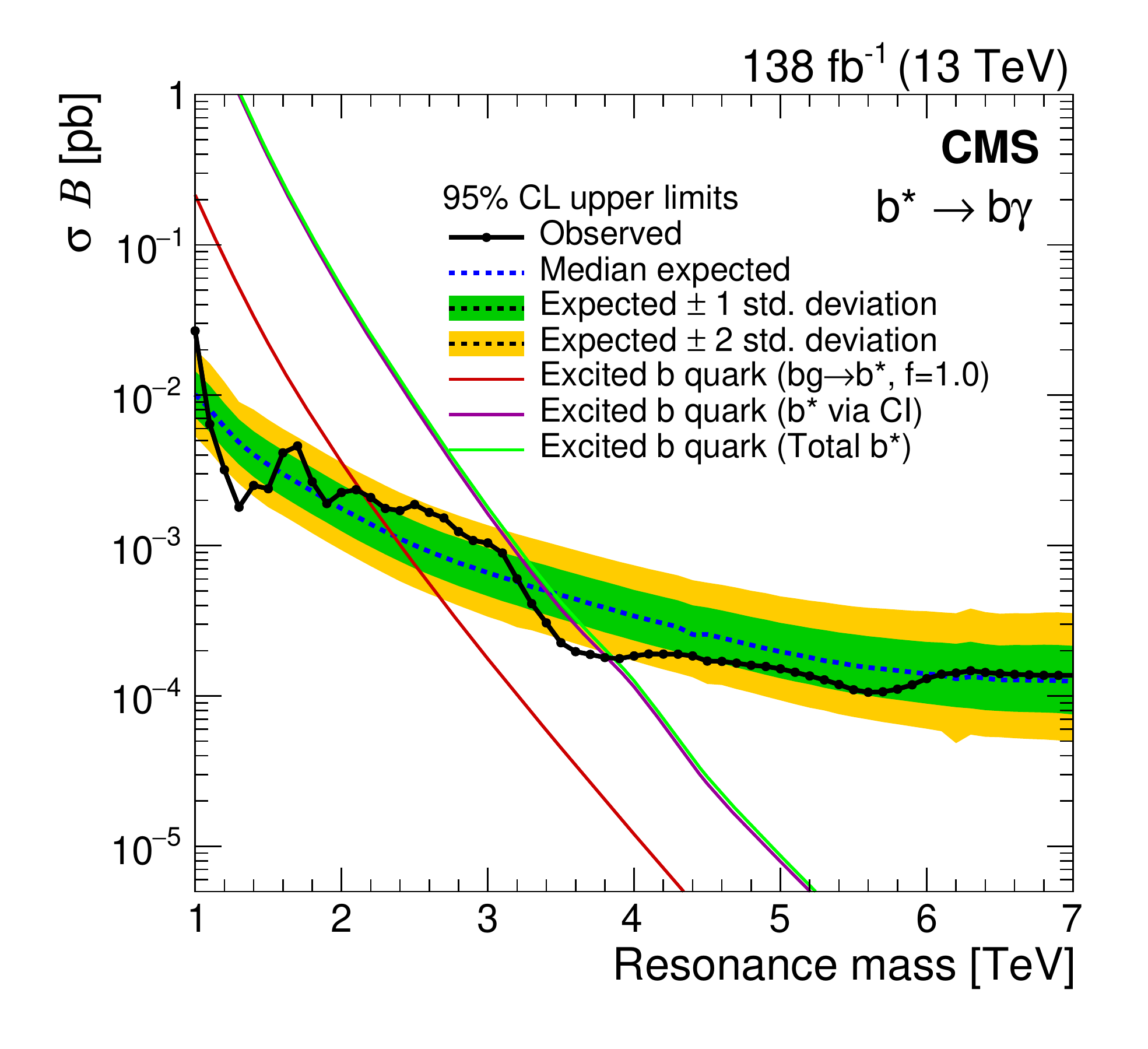}
\includegraphics[width=0.49\textwidth]{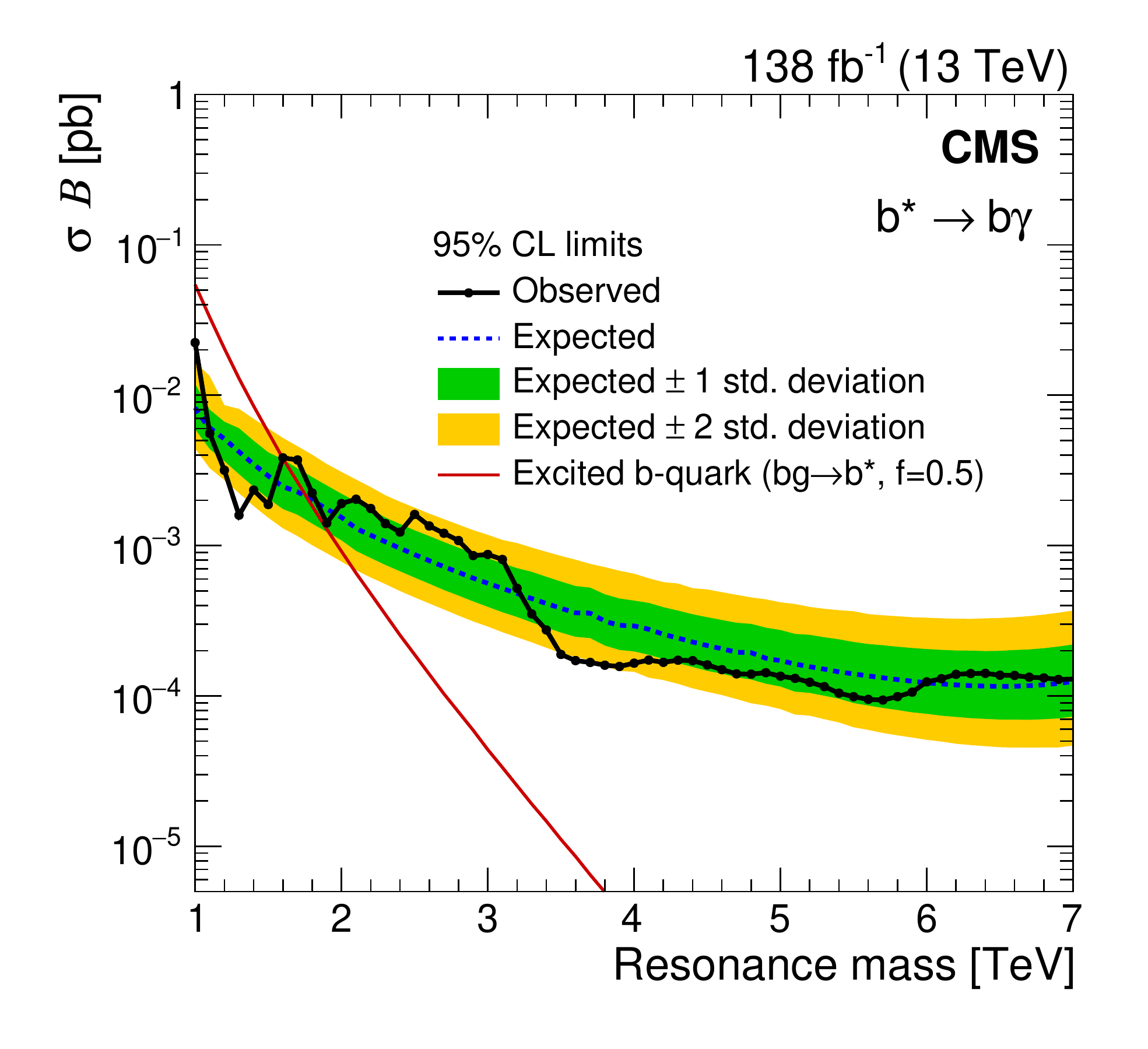}
\includegraphics[width=0.49\textwidth]{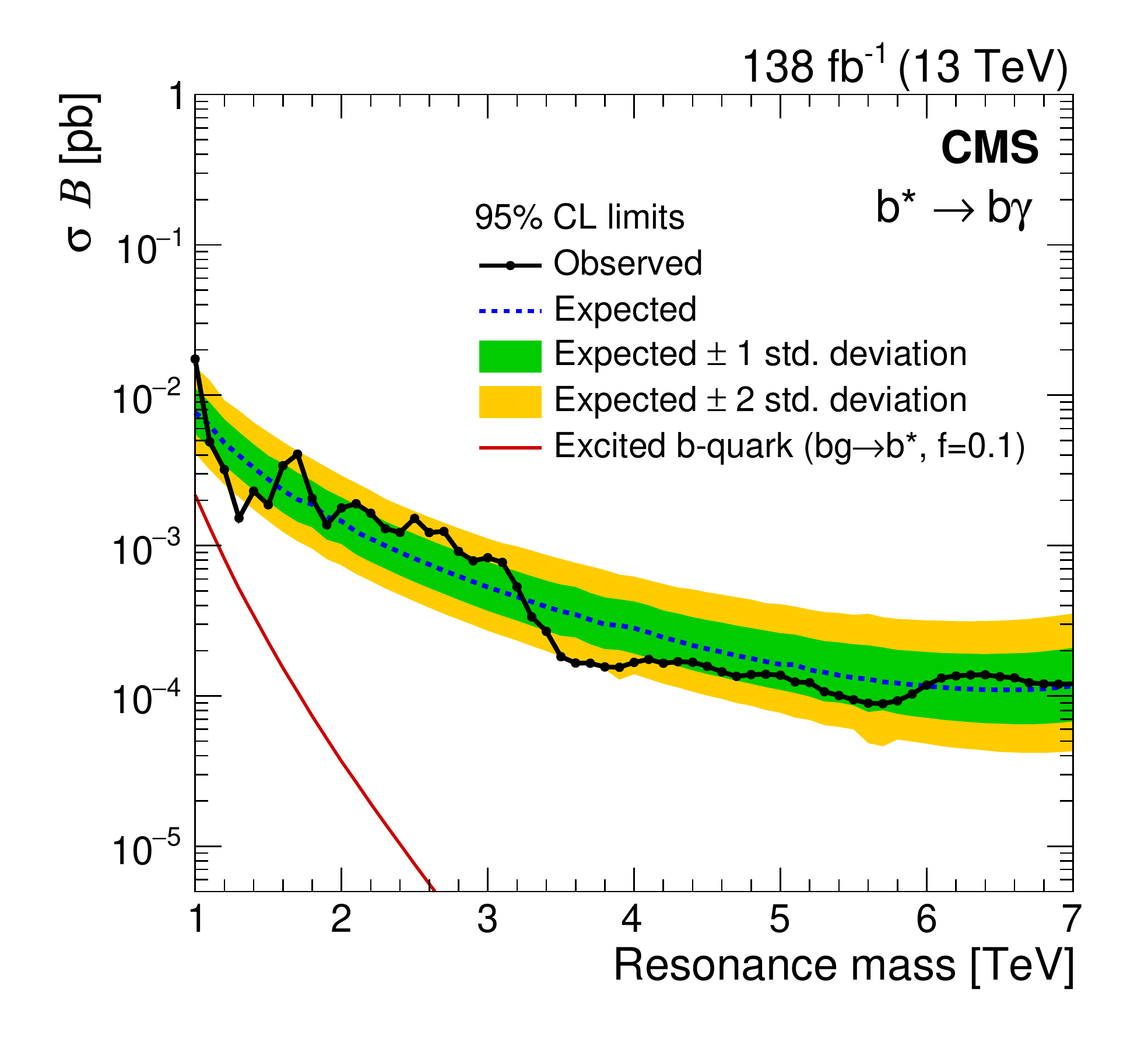}
\caption{The expected (dashed) and observed (solid) 95\% \CL upper limits on the product of the cross section and branching fraction, as functions of \PQbst mass, for coupling strength $f=1.0$ (upper left), $f=0.5$ (upper right), and $f=0.1$ (lower). The green (inner) and yellow (outer) bands correspond to 1 and 2 standard deviation uncertainties in the expected limit, respectively. The limits are compared with the theoretical predictions for \PQbst production for the corresponding coupling strengths. For coupling $f=1.0$, the limits are compared for \PQbst production by gauge interactions (red), contact interactions (violet), and total \PQbst signal by the addition of these two production modes.}
\label{fig:bstarlimit}
\end{figure}

\begin{figure}[htb!]
\centering
\includegraphics[width=0.49\textwidth]{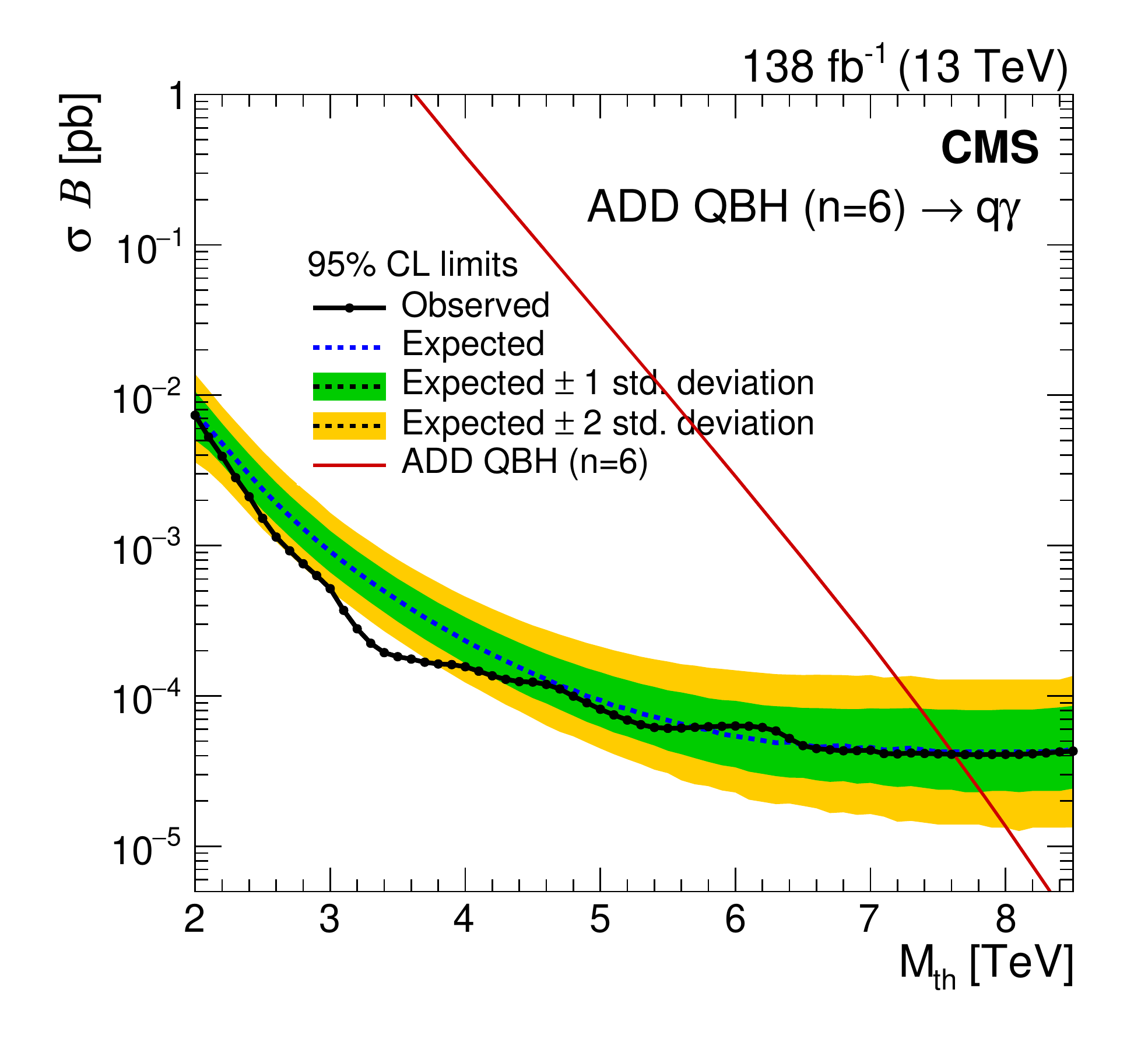}
\includegraphics[width=0.49\textwidth]{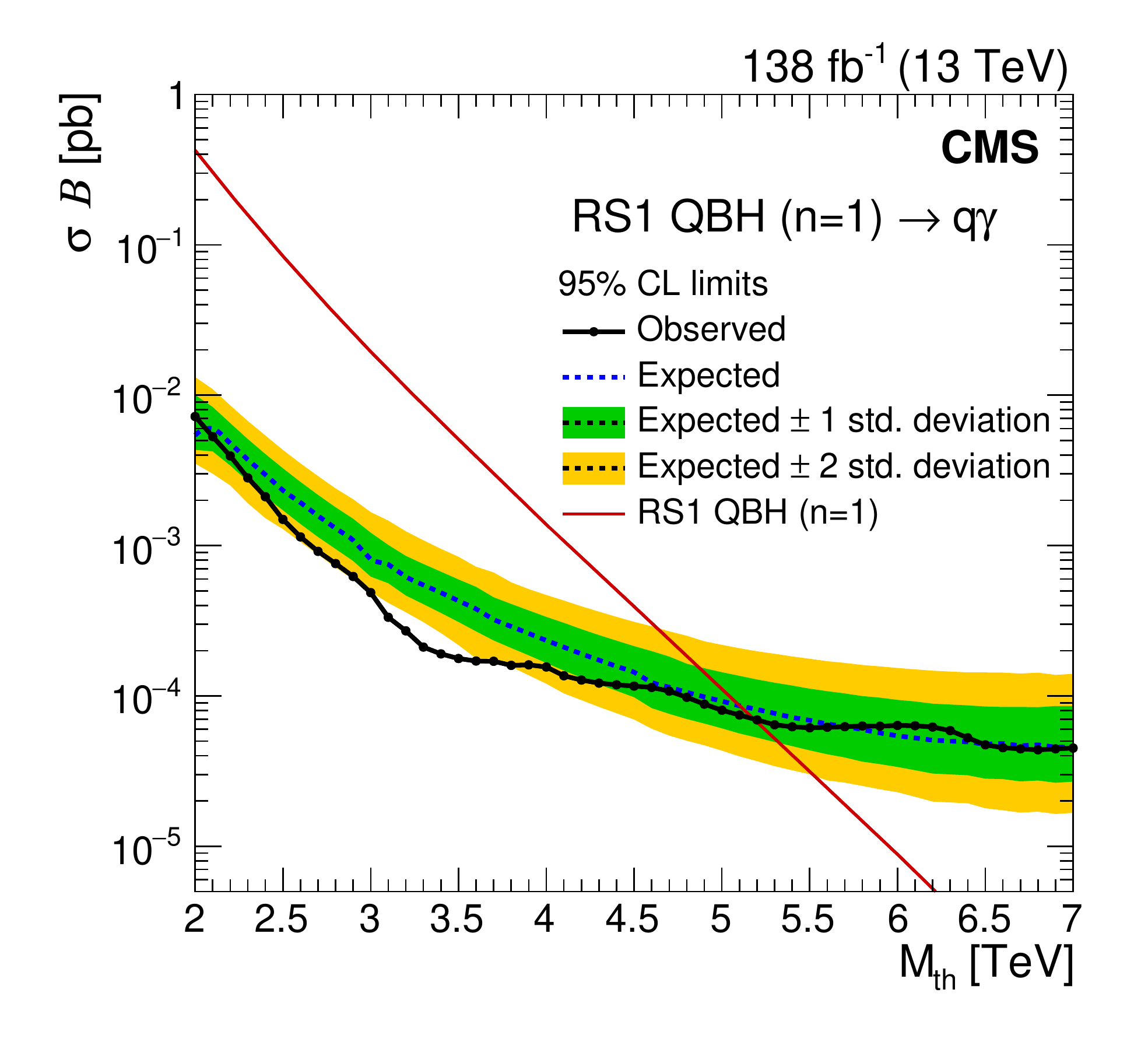}
\caption{The expected (dashed) and observed (solid) 95\% \CL upper limit on the product of the cross section and branching fraction as a function of the minimum black hole mass for the ADD ($n=6$) and RS1 ($n=1$) models on the left and right, respectively. The green (inner) and yellow (outer) bands correspond to 1 and 2 standard deviation uncertainties in the expected limit. The limits are compared with the theoretical predictions for QBH production for ADD ($n=6$) and RS1 ($n=1$) models.}
\label{fig:QBHlimit}
\end{figure}

The observed mass spectrum is found to be consistent with the background-only prediction. We set limits on potential signals by using a \CLs approach~\cite{CLS1,CLS2} based on an asymptotic approximation~\cite{ATLAS:1379837,Cowan:2010js} of the multibin counting experiment likelihood, taken to be a product of Poisson distributions corresponding to individual bins:
\begin{linenomath}
\begin{equation}
\mathcal{L}(\text{data}|\mu,\theta) = \prod_{i=1}^{n_\text{bin}}\mathrm{P}\left(x_{i}|s_{i}(\mu,\theta) + b_{i}(\theta)\right)\mathrm{C}(\theta),
\end{equation}
\end{linenomath}
where P stands for the Poisson probability to observe $x_{i}$ data events in the $i^{\text{th}}$ bin, C represents a constraint term for the nuisance parameters ($\theta$) associated with the systematic uncertainties, $\mu$ is the signal strength (parameter of interest), $s_{i}(\mu,\theta)$ the corresponding signal
yield, $b_{i}(\theta)$ the corresponding background yield, and the product runs over the number of bins ($n_\text{bin}$). The systematic uncertainties are considered as nuisance parameters with log-normal constraints on the integrated luminosity and trigger efficiency, and Gaussian constraints on the energy scales and resolutions for the photon and jet. The observed and expected limits are computed and shown in Figs.~\ref{fig:qstarlimit}--\ref{fig:QBHlimit} in terms of the product of the cross-section and the branching fraction, \sigmaBR. 

The observed (expected) lower limits on \McPqst are found to be 6.0 (6.0), 5.4 (5.2) and 2.4 (2.8)\TeV for the coupling strength $f = 1.0$, 0.5, and 0.1, respectively.

For the \PQbst resonance search with the process $\PQb\Pg \to \PQbst \to \PQb\Pgg$, the observed (expected) lower bounds are 2.2 (2.3) and 1.6 (1.7)\TeV for a coupling strength of $f=1.0$ and 0.5, respectively. For $f=0.1$, the analysis did not have sufficient sensitivity to set a mass limit. For \PQbst production via CI, several processes contribute, with the dominant one being $\qqbar\to\PAQb{\PQbst} (\text{or}\,\PQb\PAQb^{\star}) \to\PAQb\PQb\PGg$. Using the same exclusion limit for the $\PQb\Pg \to \PQbst$ signal with $f=1.0$, the observed lower bound on the \PQbst mass is 3.8\TeV when combining both the processes (\PQbst produced via gauge interaction and the resonant component of \PQbst via CI) in the total signal cross section, as shown in Fig.~\ref{fig:bstarlimit} (upper left).

For the QBH search with the ADD (RS1) model with $n=6$ (1) extra dimensions, the observed lower bound on the QBH mass is 7.5 (5.2)\TeV and the expected exclusion limit is 7.5 (5.1)\TeV. In comparison to \PQqst and \PQbst resonances, the exclusion limits are stronger for QBHs. 

The above-mentioned observed lower mass limits of 6.0 and 2.2 TeV on excited light-flavor quarks and excited bottom quarks, for the coupling strength of $f=1.0$, are improvements compared to the previous results in the same channel of 5.5 and 1.8\TeV, respectively~\cite{Sirunyan:2017fho}. The lower QBH mass limits for the ADD (RS1) models are improved from 7.1 (4.4)\TeV to 7.5 (5.2)\TeV in the same channel compared to the previous results reported in Ref.~\cite{atlas_qbh}.  

\begin{figure}[!htbp]
\centering
\includegraphics[width=0.99\textwidth]{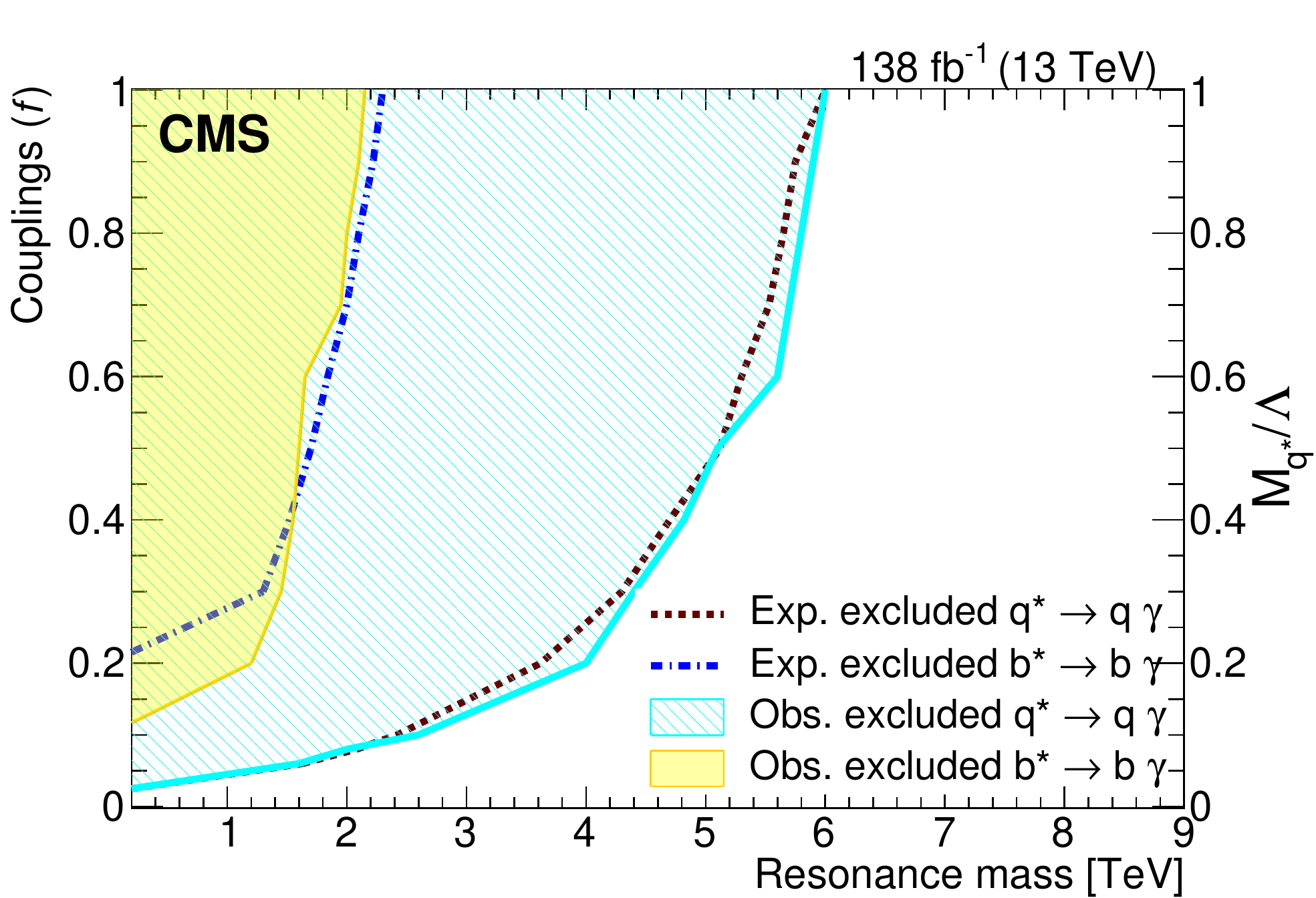}
\caption{The expected and observed 95\% \CL exclusion mass limit variation with the SM couplings for the excited \PQqst and \PQbst signal models.}
\label{fig:2Dplot}
\end{figure}

The two-dimensional representation of the excluded mass region for excited quarks (\PQqst and \PQbst) as a
function of the coupling strength $f$, with the assumption $\McPqst = \Lambda$, is shown in
Fig.~\ref{fig:2Dplot}. The exclusion of the excited quark mass, as a function of $\McPqst/\Lambda$, is also presented  in the same figure on the right hand axis. The excluded \McPqst region could also be interpreted in
terms of the $\McPqst/\Lambda$ ratio; this interpretation is based on the fact that the detector resolution effects dominate compared to the change in the signal intrinsic width ($\Gamma_{\PQqst} \sim 0.03 f^{2}\McPqst$)~\cite{Baur1989kv} arising from the change in the coupling strength $f$. Consequently, the variation of $f$ effectively changes the production cross section in a similar way to the variations in $\McPqst/\Lambda$, since both appear as multiplicative factors in the cross section calculation. 

\section {Summary}
\label{sec:summary}
A search for resonances in the \gammajet final state has been performed using proton-proton collision data at
\sqrtsXIII, corresponding to an integrated luminosity of 138\fbinv. Compared to previous searches in this final state, this analysis features improved \PQb jet tagging efficiency and signal mass resolution, as well as a significantly larger data set. The results are compared with standard model extensions postulating excited quarks, excited \PQb quarks, and quantum black holes.  

For the coupling strength $f = 1.0$, the observed (expected) lower mass bounds on the excited light-flavor quarks and excited bottom quarks from gauge interactions are estimated to be 6.0 (6.0) and 2.2 (2.3)\TeV, respectively. For the same coupling strength, the observed lower mass bounds on the excited bottom quarks extends to 3.8\TeV, when the production mode from contact interactions is also considered. For quantum black hole production, observed exclusion limits  are estimated to be 7.5 and 5.2\TeV for the Arkani-Hamed--Dimopoulos--Dvali and the Randall--Sundrum  models with 6 and 1 extra dimensions, respectively. These lower mass bounds are extended compared to previous results by 0.4--0.8\TeV~\cite{Sirunyan:2017fho,atlas_qbh} and are the most stringent to date among those obtained in the \gammajet final state. 

\begin{acknowledgments}
We congratulate our colleagues in the CERN accelerator departments for the excellent performance of the LHC and thank the technical and administrative staffs at CERN and at other CMS institutes for their contributions to the success of the CMS effort. In addition, we gratefully acknowledge the computing centers and personnel of the Worldwide LHC Computing Grid and other centers for delivering so effectively the computing infrastructure essential to our analyses. Finally, we acknowledge the enduring support for the construction and operation of the LHC, the CMS detector, and the supporting computing infrastructure provided by the following funding agencies: BMBWF and FWF (Austria); FNRS and FWO (Belgium); CNPq, CAPES, FAPERJ, FAPERGS, and FAPESP (Brazil); MES and BNSF (Bulgaria); CERN; CAS, MoST, and NSFC (China); MINCIENCIAS (Colombia); MSES and CSF (Croatia); RIF (Cyprus); SENESCYT (Ecuador); MoER, ERC PUT and ERDF (Estonia); Academy of Finland, MEC, and HIP (Finland); CEA and CNRS/IN2P3 (France); BMBF, DFG, and HGF (Germany); GSRI (Greece); NKFIH (Hungary); DAE and DST (India); IPM (Iran); SFI (Ireland); INFN (Italy); MSIP and NRF (Republic of Korea); MES (Latvia); LAS (Lithuania); MOE and UM (Malaysia); BUAP, CINVESTAV, CONACYT, LNS, SEP, and UASLP-FAI (Mexico); MOS (Montenegro); MBIE (New Zealand); PAEC (Pakistan); MES and NSC (Poland); FCT (Portugal);  MESTD (Serbia); MCIN/AEI and PCTI (Spain); MOSTR (Sri Lanka); Swiss Funding Agencies (Switzerland); MST (Taipei); MHESI and NSTDA (Thailand); TUBITAK and TENMAK (Turkey); NASU (Ukraine); STFC (United Kingdom); DOE and NSF (USA).
  
  \hyphenation{Rachada-pisek} Individuals have received support from the Marie-Curie program and the European Research Council and Horizon 2020 Grant, contract Nos.\ 675440, 724704, 752730, 758316, 765710, 824093, 884104, and COST Action CA16108 (European Union); the Leventis Foundation; the Alfred P.\ Sloan Foundation; the Alexander von Humboldt Foundation; the Belgian Federal Science Policy Office; the Fonds pour la Formation \`a la Recherche dans l'Industrie et dans l'Agriculture (FRIA-Belgium); the Agentschap voor Innovatie door Wetenschap en Technologie (IWT-Belgium); the F.R.S.-FNRS and FWO (Belgium) under the ``Excellence of Science -- EOS" -- be.h project n.\ 30820817; the Beijing Municipal Science \& Technology Commission, No. Z191100007219010; the Ministry of Education, Youth and Sports (MEYS) of the Czech Republic; the Hellenic Foundation for Research and Innovation (HFRI), Project Number 2288 (Greece); the Deutsche Forschungsgemeinschaft (DFG), under Germany's Excellence Strategy -- EXC 2121 ``Quantum Universe" -- 390833306, and under project number 400140256 - GRK2497; the Hungarian Academy of Sciences, the New National Excellence Program - \'UNKP, the NKFIH research grants K 124845, K 124850, K 128713, K 128786, K 129058, K 131991, K 133046, K 138136, K 143460, K 143477, 2020-2.2.1-ED-2021-00181, and TKP2021-NKTA-64 (Hungary); the Council of Science and Industrial Research, and University Grants Commission, India; the Latvian Council of Science; the Ministry of Education and Science, project no. 2022/WK/14, and the National Science Center, contracts Opus 2021/41/B/ST2/01369 and 2021/43/B/ST2/01552 (Poland); the Funda\c{c}\~ao para a Ci\^encia e a Tecnologia, grant CEECIND/01334/2018 (Portugal); the National Priorities Research Program by Qatar National Research Fund; MCIN/AEI/10.13039/501100011033, ERDF ``a way of making Europe", and the Programa Estatal de Fomento de la Investigaci{\'o}n Cient{\'i}fica y T{\'e}cnica de Excelencia Mar\'{\i}a de Maeztu, grant MDM-2017-0765 and Programa Severo Ochoa del Principado de Asturias (Spain); the Chulalongkorn Academic into Its 2nd Century Project Advancement Project, and the National Science, Research and Innovation Fund via the Program Management Unit for Human Resources \& Institutional Development, Research and Innovation, grant B05F650021 (Thailand); the Kavli Foundation; the Nvidia Corporation; the SuperMicro Corporation; the Welch Foundation, contract C-1845; and the Weston Havens Foundation (USA).

\end{acknowledgments}
\bibliography{auto_generated}
\cleardoublepage \appendix\section{The CMS Collaboration \label{app:collab}}\begin{sloppypar}\hyphenpenalty=5000\widowpenalty=500\clubpenalty=5000
\cmsinstitute{Yerevan Physics Institute, Yerevan, Armenia}
{\tolerance=6000
A.~Tumasyan\cmsAuthorMark{1}\cmsorcid{0009-0000-0684-6742}
\par}
\cmsinstitute{Institut f\"{u}r Hochenergiephysik, Vienna, Austria}
{\tolerance=6000
W.~Adam\cmsorcid{0000-0001-9099-4341}, J.W.~Andrejkovic, T.~Bergauer\cmsorcid{0000-0002-5786-0293}, S.~Chatterjee\cmsorcid{0000-0003-2660-0349}, K.~Damanakis\cmsorcid{0000-0001-5389-2872}, M.~Dragicevic\cmsorcid{0000-0003-1967-6783}, A.~Escalante~Del~Valle\cmsorcid{0000-0002-9702-6359}, P.S.~Hussain\cmsorcid{0000-0002-4825-5278}, M.~Jeitler\cmsAuthorMark{2}\cmsorcid{0000-0002-5141-9560}, N.~Krammer\cmsorcid{0000-0002-0548-0985}, L.~Lechner\cmsorcid{0000-0002-3065-1141}, D.~Liko\cmsorcid{0000-0002-3380-473X}, I.~Mikulec\cmsorcid{0000-0003-0385-2746}, P.~Paulitsch, J.~Schieck\cmsAuthorMark{2}\cmsorcid{0000-0002-1058-8093}, R.~Sch\"{o}fbeck\cmsorcid{0000-0002-2332-8784}, D.~Schwarz\cmsorcid{0000-0002-3821-7331}, M.~Sonawane\cmsorcid{0000-0003-0510-7010}, S.~Templ\cmsorcid{0000-0003-3137-5692}, W.~Waltenberger\cmsorcid{0000-0002-6215-7228}, C.-E.~Wulz\cmsAuthorMark{2}\cmsorcid{0000-0001-9226-5812}
% ...
\par}
\vskip\cmsinstskip
\dag:~Deceased\\
$^{1}$Also at Yerevan State University, Yerevan, Armenia\\
$^{2}$Also at TU Wien, Vienna, Austria\\
% ...
\end{sloppypar}
%%% END EDITABLE REGION %%%
% skeleton_end
\end{document}